%% file: Stable-fp-2023-nov18-arxiv.tex
\documentclass[12pt]{article}

\input{packages-v2}

\usepackage{helvet}
\usepackage{graphicx,amssymb,amsmath,amsfonts,amsthm,color,url, placeins}
\usepackage{wrapfig}
\usepackage{tabstackengine}

\renewcommand\section{\@startsection{section}{1}{\z@}%
                                   {-3.0ex \@plus -1ex \@minus -.2ex}%
                                   {1.5ex \@plus.2ex}%
                                   {\normalfont\sffamily\large\bfseries}}
\renewcommand\subsection{\@startsection{subsection}{2}{\z@}%
                                     {-2.75ex\@plus -1ex \@minus -.2ex}%
                                     {1.5ex \@plus .2ex}%
                                   {\normalfont\sffamily\large}}
\renewcommand\subsubsection{\@startsection{subsubsection}{3}{\z@}%
                                     {-2.75ex\@plus -1ex \@minus -.2ex}%
                                     {1.5ex \@plus .2ex}%
                                   {\normalfont\sffamily\large}}

\newcommand{\od}{\stackrel{\mbox {\tiny {def}}}{=}}

\def\RR{\mathbb{R}}

\def\d{\mathrm{d}}

\def\RR{\mathbb{R}}

\def\RR{\mathbb{R}}

\def\det{\operatorname{det}}

\def\supp{\operatorname{supp}}

\def\od{\stackrel{\mathrm{def}}{=}}

\def\supp{\operatorname{supp}}

\def\sgn{\operatorname{sgn}}
\def\FP{\operatorname{FP}}

\def\idx{\operatorname{idx}}

\def\Gtil{\widetilde{G}}
\def\Ghat{\widehat{G}}
\def\eig{\operatorname{eig}}

\def\Tr{\operatorname{Tr}}

\usepackage{color}
\definecolor{cherry}{rgb}{0.9,.1,.2}

\begin{document}

\noindent {\Large \bf Stable fixed points of combinatorial threshold-linear networks}\\
Carina Curto, Jesse Geneson, Katherine Morrison\\
\medskip 
November 18, 2023

\noindent {\bf Abstract.} 
Combinatorial threshold-linear networks (CTLNs) are a special class of recurrent neural networks whose dynamics are tightly controlled by an underlying directed graph.  Recurrent networks have long been used as models for associative memory and pattern completion, with stable fixed points playing the role of stored memory patterns in the network.  In prior work, we showed that \emph{target-free cliques} of the graph correspond to stable fixed points of the dynamics, and we conjectured that these are the only stable fixed points possible \cite{CTLN-preprint, fp-paper}.  In this paper, we prove that the conjecture holds in a variety of special cases, including for networks with very strong inhibition and graphs of size $n \leq 4$. We also provide further evidence for the conjecture by showing that sparse graphs and graphs that are nearly cliques can never support stable fixed points.
Finally, we translate some results from extremal combinatorics to obtain an upper bound on the number of stable fixed points of CTLNs in cases where the conjecture holds.  \\


\tableofcontents

\medskip 

\section{Introduction}

One of the major challenges in theoretical neuroscience is to relate network connectivity to function. In the study of attractor neural networks, dating back at least to the Hopfield model \cite{Hopfield1, Amit-ANNs}, the attractors of interest have typically been stable fixed points of the dynamics. These fixed points represent stored memory patterns, and the function of the network is to transform an input pattern into a stable fixed point via the evolution of network dynamics. In this way, attractor neural networks can perform pattern completion, image classification, and memory retrieval \cite{XieHahnSeung, pattern-completion, Hillar2014}. Stable fixed points have also been used to model sensory representations, such as orientation selectivity and surround suppression in visual cortex \cite{FersterMiller, MillerFerster}, as well as position coding in hippocampus \cite{Tsodyks1999}, perceptual bistability \cite{Rinzel-perceptual-bistability}, and decision-making \cite{Brody2017}. How does a network encode a set of attractors? More specifically, given a network connectivity matrix $W$, what can we say about the stable fixed points? 

In this work, we address this question in the context of combinatorial threshold-linear networks (CTLNs). These are a special case of threshold-linear networks (TLNs), which are recurrent network models that have been commonly used in computational neuroscience for decades \cite{AppendixE, TrevesRolls, Seung-Nature, XieHahnSeung, HahnSeungSlotine, flex-memory, net-encoding, pattern-completion, Fitzgerald2020}.  In the case of CTLNs, the connectivity matrix, $W$, is tightly controlled by a directed graph, $G$, leading to a fixed point structure that is closely related to properties of $G$. In prior work, we showed that certain cliques of $G$, called \emph{target-free cliques}, correspond to stable fixed points of the dynamics. Furthermore, we conjectured that these are the only stable fixed points possible \cite{CTLN-preprint, fp-paper}.  Here we prove that the conjecture holds in a variety of special cases, including for networks with very strong inhibition and graphs of size $n \leq 4$. We also provide further evidence for the conjecture by showing that sparse graphs and graphs that are nearly cliques can never support stable fixed points.
Finally, we translate some results from extremal combinatorics to obtain an upper bound on the number of stable fixed points of CTLNs in cases where the conjecture holds. 

We begin with some basic background on TLNs and CTLNs, enough so that we can state the main results.

\subsection{Threshold-linear networks}

Briefly, a TLN is a rate model consisting of $n$ nodes, with dynamics governed by the system of ordinary differential equations:
\begin{equation}\label{eq:dynamics}
\dfrac{dx_i}{dt} = -x_i + \left[\sum_{j=1}^n W_{ij}x_j+b_i \right]_+, \quad i = 1,\ldots,n.
\end{equation}
The dynamic variables $x_1,\ldots,x_n$ give the activity levels\footnote{If the nodes are neurons, the activity level is typically called a `firing rate.'} of nodes $1,\ldots,n$.
The matrix entries $W_{ij}$ are directed connection strengths between pairs of nodes, the vector $b=(b_1,\ldots,b_n) \in \RR^n$ represents the external drive to each node, and the threshold-nonlinearity $[\cdot]_+$ is given by $[y]_+ = \max\{y,0\}$. 

TLNs are a particularly appealing model because they exhibit the full range of nonlinear behaviors, such as multistability, periodic attractors, and chaos, while remaining relatively mathematically tractable because of their piecewise linear dynamics.  \emph{Multistability}, the coexistence of multiple stable fixed points in a network, has made TLNs particularly useful as models of associative memory storage and retrieval, similar to the Hopfield model \cite{Hopfield1}.  In this context, the stable fixed points represent the set of memories encoded in the network.  The network can then perform a sort of \emph{pattern completion}: the activity evolves from a given initial condition to converge to one of the multiple stored patterns of the network.  The first major developments in the mathematical theory of TLNs were made for \emph{symmetric networks}, with a focus on characterizing the stable fixed points \cite{Seung-Nature, XieHahnSeung,  HahnSeungSlotine}.  

\subsection{Fixed points of TLNs}\label{sec:general-TLNs}
Recall that a fixed point is a point $x^* \in \RR^n$ that satisfies $dx_i/dt|_{x=x^*} = 0$ for each $i \in \{1, \ldots, n\}$.  The  {\it support} of a fixed point is the subset of active nodes, 
$$\supp{x^*} \od \{i \mid x^*_i>0\}.$$   We restrict to considering TLNs that are \emph{nondegenerate} (see Section~\ref{sec:general-background} for the precise definition).  One consequence of the nondegeneracy condition (that is generically satisfied) is that there is at most one fixed point per support \cite{fp-paper}.  We can thus label all the fixed points of a given network by their supports. We denote this as:
\begin{eqnarray*}
\FP(W,b) &\od& \{\sigma \subseteq [n] \mid   \sigma = \supp{x^*} \text{ for some } \text{fixed pt } x^* \text{ of the associated TLN} \},
\end{eqnarray*}
where $[n] \od \{1,\ldots,n\}$.  Note that for each support $\sigma \in \FP(W,b)$, the fixed point itself is easily recovered. Outside the support, $x_i^* = 0$ for all $i \not\in \sigma$. Within the support, $x^*$ is given by:
\begin{equation}\label{eq:x_sigma^*}
x_\sigma^* = (I-W_\sigma)^{-1} b_\sigma,
\end{equation}
where $x_\sigma^*$ and $b_\sigma$ are column vectors obtained by restricting $x^*$ and $b$ to the indices in $\sigma$, and $W_\sigma$ is the induced principal submatrix obtained by restricting rows and columns of $W$ to $\sigma$.

Recall that a fixed point $x^*$ is \emph{stable} if small perturbations of a solution from that point decay over time; consequently, within a neighborhood of the point, all trajectories converge to that fixed point.  Otherwise, the fixed point is said to be \emph{unstable}.  The following result shows that stability can be checked directly from the matrix $W$; this was first proven for symmetric $W$ in \cite{HahnSeungSlotine}, and later generalized to arbitrary $W$ in \cite{flex-memory}.  

\begin{theorem}[\cite{HahnSeungSlotine, flex-memory}] \label{thm:stable-TLN}
Let $(W,b)$ be a TLN on $n$ nodes, and consider $\sigma \in \FP(W,b)$.   Then the fixed point of \eqref{eq:dynamics} with support $\sigma$ is stable if and only if the principal submatrix $(-I+W)_\sigma$ is a stable matrix -- i.e., if all its eigenvalues have negative real part.
\end{theorem}

Significant work has been done characterizing the collection of stable fixed point supports of symmetric $W$ through the lens of \emph{permitted sets}, initially in \cite{HahnSeungSlotine} and extended in \cite{flex-memory, net-encoding, pattern-completion}.  
These advances were largely confined to the case of symmetric $W$ because there is more machinery for analyzing eigenvalues in that setting, such as Cauchy's Interlacing Theorem and results on nondegenerate square distance matrices.  Thus, to make progress in the nonsymmetric case, it is useful to restrict to a different class of networks that extend well beyond the symmetric case, but maintain mathematical tractability.

\subsection{The CTLN model and stable fixed points of CTLNs} \label{sec:CTLN-setup}
\begin{figure}[!ht]
\begin{center}
\includegraphics[width=6in]{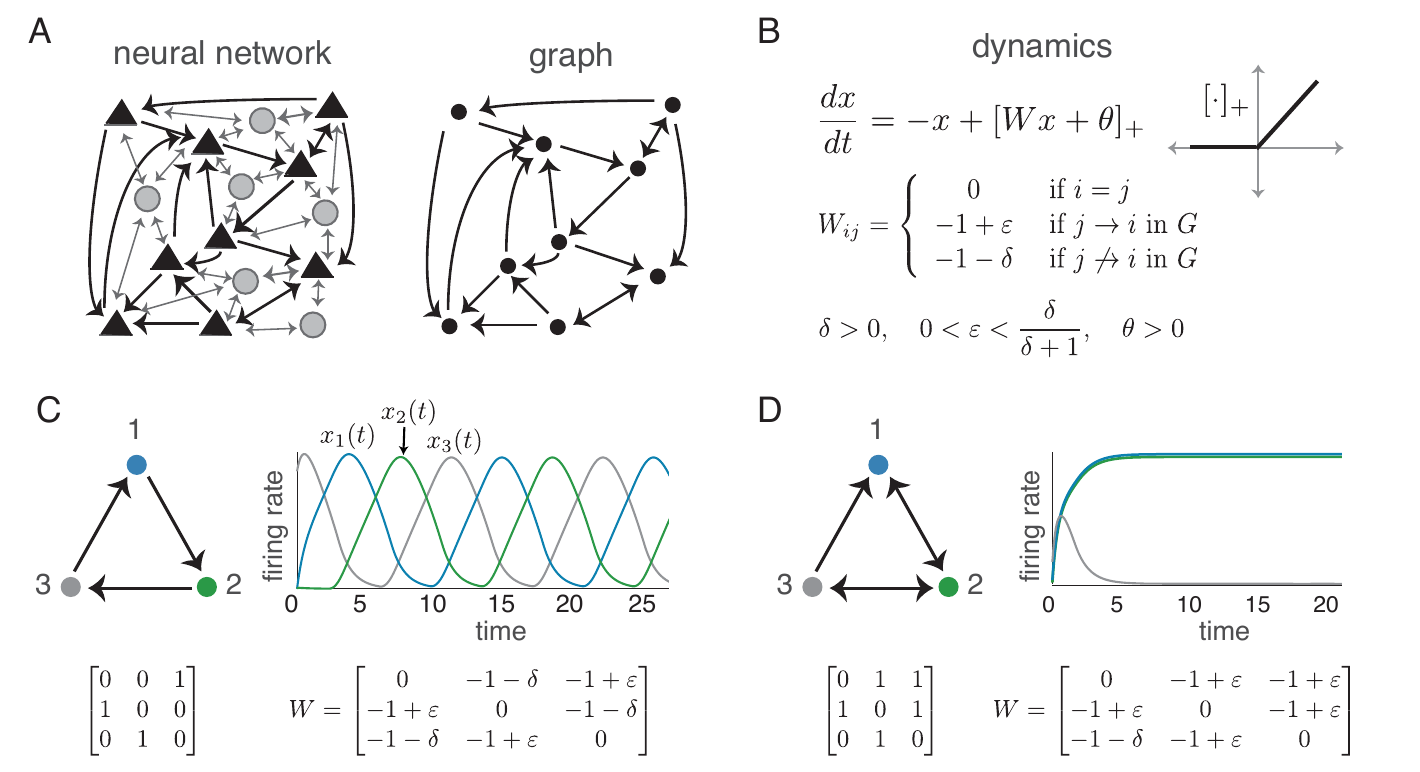}
\vspace{.05in}
\caption{{\bf Combinatorial threshold-linear networks with strong and weak inhibition} (modeling excitatory neurons in a sea of inhibition). (A) (Left) A neural network with excitatory pyramidal neurons (triangles) and a background network of inhibitory interneurons (gray circles) that produce a global inhibition.   (Right) The graph of the network retains only the excitatory neurons and the connections between them.  In the corresponding CTLN, the arrows in the graph  indicate weak inhibition from the sum of excitation and a global background inhibition.  The absence of an edge thus indicates strong inhibition. (B) The equations for a CTLN network.
(C) (Left) The 3-cycle graph with its corresponding adjacency matrix and $W$ matrix below.  (Right) Network activity follows the arrows in the graph, with peak activity occurring sequentially in the cyclic order 123.  (D) A CTLN with one stable fixed point, which has support $\{1,2\}$ (network activity shown on right).  Note that $\{1,2\}$ is a {\it target-free} clique. The clique $\{2,3\}$ does not have a corresponding fixed point; node $1$ is a target of this clique.  All simulations have parameters $\varepsilon=0.25$, $\delta=0.5.$, and $\theta=1$.}
\label{fig:network-setup}
\end{center}
\vspace{-.25in}
\end{figure}

Combinatorial threshold-linear networks (CTLNs) are a special class of threshold-linear networks, first introduced in \cite{CTLN-preprint}, whose dynamics are tightly controlled by an underlying directed graph.  CTLNs have uniform inputs, $b_i=\theta$, and a connectivity matrix $W$ that is fully determined by a simple\footnote{A directed graph is \emph{simple} if it does not have self-loops or multiple edges between a pair of nodes} directed graph $G$ and continuous parameters $\varepsilon$ and $\delta$. Specifically $W = W(G,\varepsilon,\delta)$ has the form:
\vspace{-.075in}
\begin{equation} \label{eq:binary-synapse}
\vspace{-.025in}
W_{ij} = \left\{\begin{array}{cl} 0 & \text{ if } i = j, \\ -1 + \varepsilon & \text{ if } j \rightarrow i \text{ in } G,\\ -1 -\delta & \text{ if } j \not\rightarrow i \text{ in } G. \end{array}\right. \quad \quad \quad \quad
\end{equation}
where $ j \rightarrow i$ indicates the presence of an edge from $j$ to $i$ in the graph $G$, while $j \not\rightarrow i$ indicates the absence of such an edge.  We additionally require that $\theta > 0$,\; $\delta >0,$ and $0 < \varepsilon < \frac{\delta}{\delta+1}$; or equivalently, $0 < \varepsilon < 1$ and $\delta> \frac{\varepsilon}{1-\varepsilon}$.  When these conditions are met, we say that the parameters are within the \emph{legal range}\footnote{Within this parameter range, $W_{ij} \leq 0$ which guarantees that the network is competitive and activity is bounded.  The particular constraint on $\varepsilon$ was motivated by a result in \cite{CTLN-preprint} to guarantee that a unidirectional edge is always unstable.} (see Figure~\ref{fig:network-setup}B).  

The simplicity of binary synapses has made the CTLN model particularly mathematically tractable, while enabling us to explore networks beyond the symmetric case.  Moreover, the tight connection between the connectivity graph $G$ and the matrix $W$ has allowed us to isolate the role of connectivity in shaping dynamics.  This has led to the development of a series of \emph{graph rules} that directly connect fixed points of the network to particular features or subgraphs of $G$ \cite{CTLN-preprint, fp-paper, sequential-att-paper, nerve-thm-paper}.  
The primary focus of this paper is a graphical characterization of the stable fixed points of CTLNs.  This requires some terminology:  A \emph{clique} is subgraph in which there are all-to-all bidirectional connections.  We say that a node $k \notin \sigma$ is a \emph{target} of $\sigma$ if $i \to k$ for every $i \in \sigma$.  We refer to cliques that do not have any targets in $G$ as \emph{target-free cliques}.

In \cite{CTLN-preprint}, it was shown that every target-free clique of $G$ is the support of a stable fixed point of any CTLN with graph $G$, irrespective of parameters (within the legal range).  Moreover, these appeared to be the \underline{only} graph structures that gave rise to stable fixed points.  These observations led to the following conjecture.

\begin{conjecture}[\cite{CTLN-preprint}]\label{conjecture}
Let \;$W=W(G, \varepsilon, \delta)$ be a CTLN on $n$ nodes with graph $G$, and let $\sigma \subseteq [n]$.  Then
$\sigma$ is the support of a stable fixed point if and only if $\sigma$ is a target-free clique. 
\end{conjecture}

\begin{figure}[!ht]
\vspace{-.05in}
\begin{center}
\includegraphics[width=6.25in]{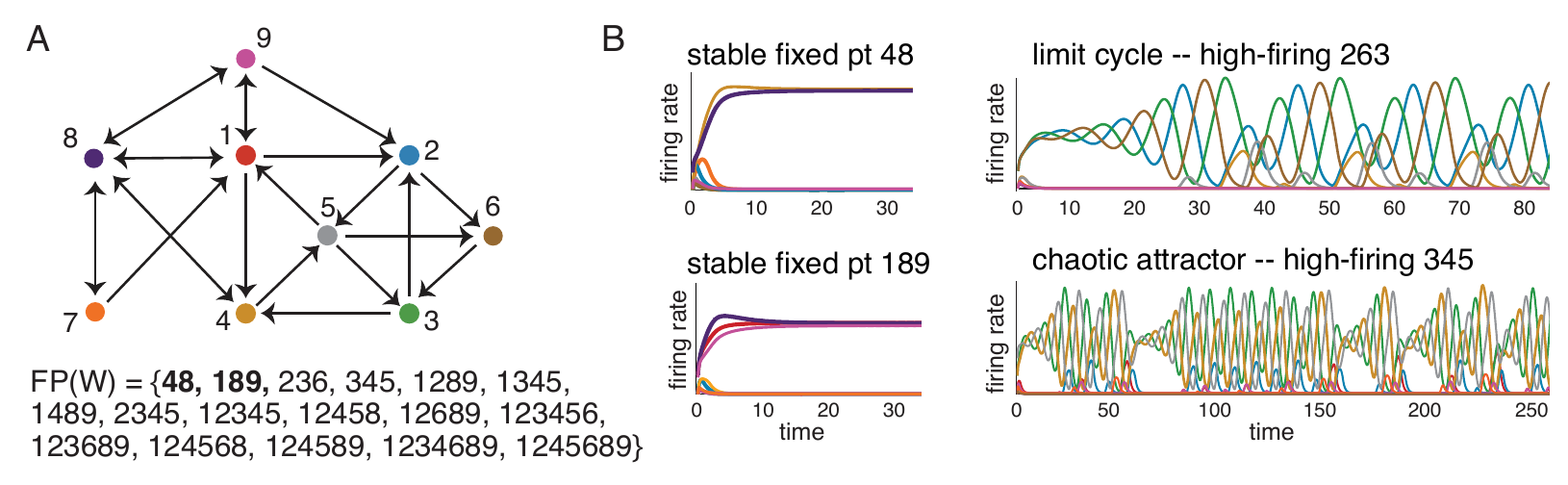}
\vspace{.05in}
\caption{{\bf A CTLN with multiple fixed points and attractors.} (A) Graph of a CTLN together with its fixed point supports $\FP(W) \od \FP(W(G, \varepsilon, \delta), \theta)$ for $\varepsilon =0.25,~\delta=0.5,~\theta=1$.  The supports of stable fixed points are bolded.  (B) The network has four attractors: two stable fixed points, one limit cycle, and one chaotic attractor.  The equations for the dynamics are identical in each case; only the initial conditions differ, and these determine which attractor the solution converges to.}
\label{fig:coexistence}
\end{center}
\vspace{-.3in}
\end{figure}

As an illustration of the conjecture, consider the CTLN in Figure~\ref{fig:coexistence}.  The collection of fixed point supports, $\FP(W)$, are given below the graph of the network, with the stable fixed points in bold.\footnote{For compactness, we write $ijk$ to denote the subset $\{i,j,k\}$.}  Note that the supports of the stable fixed points are precisely the target-free cliques $48$ and $189$.  The other cliques in the graph all have targets and consequently do not support fixed points of the network (see Theorem~\ref{thm:cliques-evals}); for example, the clique $78$ has node $1$ as a target, and $78 \notin \FP(W)$.  Also, notice that no other type of subgraph supports a stable fixed point other than the target-free cliques.  

Figure~\ref{fig:coexistence}B shows the attractors of the CTLN, which include two dynamic attractors in addition to the stable fixed point attractors noted above.  Note that the high-firing nodes in each of these dynamic attractors correspond to the unstable fixed points with minimal supports $236$ and $345$.  These minimal fixed points come from special types of subnetworks, known as \emph{core motifs}, which have been observed computationally to give rise to attractors of CTLNs \cite{core-motifs}.  Target-free cliques are provably core motifs, and we conjecture these are the only core motifs that correspond to stable fixed points.

The remainder of this paper is focused on collecting evidence towards proving Conjecture~\ref{conjecture}.  To state these results, we first review some useful graph-theoretic terminology.

\paragraph{Graph theory terminology.} Let $G$ be a simple directed graph on $n$ nodes and $\sigma \subseteq [n]$.  We denote by $G|_\sigma$ the induced subgraph obtained by restricting to the nodes in $\sigma$ and keeping only the edges between nodes in $\sigma$.  Recall that a subgraph $G|_\sigma$ is a \emph{clique} if every pair of nodes in $\sigma$ has a bidirectional edge between them.  As an abuse of notation, when $G$ is given, we may refer to a subset $\sigma$ as being a clique to indicate that $G|_\sigma$ is a clique.   A clique is called \emph{maximal} if it is not properly contained in any larger clique.  Note that every \emph{target-free clique} must be maximal, since any larger clique containing it would have a node $k \notin \sigma$ that is a target of $\sigma$.  
As an illustration of these terms, consider the graph in Figure~\ref{fig:coexistence}A.  We see that $189$ is a target-free clique, and thus is necessarily maximal.  The subsets $18$, $19$, and $89$ are all non-maximal cliques, and thus have targets.  Finally, $78$ is a maximal clique that is not target-free: node $1$ is a target.

As a generalization of cliques, we also consider directed cliques.  We say that $G|_\sigma$, or equivalently $\sigma$, is a \emph{directed clique} if there exists an ordering of the nodes $1, \ldots, |\sigma|$ such that $i \to j$ in $G|_\sigma$ whenever $i < j$.  Note that there are no constraints on the back edges $j \to i$ when $i <j$, and thus directed cliques include cliques as a special case.  In Figure~\ref{fig:coexistence}A, the subgraphs $G|_{256}$, $G|_{129}$, and $G|_{148}$ are all examples of directed cliques.  

A graph is \emph{oriented} if it has \underline{no} bidirectional edges, and thus no cliques of size greater than one.  Figure~\ref{fig:network-setup}C shows an example oriented graph, the 3-cycle.  As another example, for the graph $G$ in Figure~\ref{fig:coexistence}A, $G|_{\{1,\ldots, 6\}}$ is oriented.  A \emph{sink} is a node with no outgoing edges in $G$.  Note that singletons are trivially cliques, and sinks are precisely the target-free cliques of size $1$.  

Finally, we will also be interested in uniform in-degree graphs.  We say that $G|_\sigma$, or equivalently $\sigma$, is \emph{uniform in-degree} if all nodes in the graph $G|_\sigma$ have the same number of incoming edges from nodes in $\sigma$.    Note that the in-degrees need not agree in the full graph $G$, only within the restricted subgraph $G|_\sigma$.  There are no constraints on the out-degrees.

\subsection{Early evidence: conjecture holds for symmetric graphs and oriented graphs}

\noindent Conjecture~\ref{conjecture} was motivated in part by the fact that it had previously been proven to hold in two extreme cases: when the graph $G$ is \emph{symmetric}\footnote{We say that a directed graph is \emph{symmetric} if between any pair of nodes there is either a bidirectional edge or no edge.}, so that it only contains bidirectional edges, and when $G$ is oriented, so that it has no bidirectional edges.  

\begin{theorem}[\cite{pattern-completion}]\label{thm:symmetric}
Let $G$ be a symmetric graph.  Then for any CTLN with graph $G$, the supports of the stable fixed points are precisely the maximal cliques of $G$.
\end{theorem}

This theorem establishes that Conjecture~\ref{conjecture} holds for symmetric $G$ because in a symmetric graph, the target-free cliques are precisely the maximal cliques. To see this, recall that in any graph a target-free clique is necessarily maximal. On the other hand, maximal cliques in symmetric graphs must be target-free because any target $k \notin \sigma$ receiving from all nodes in $\sigma$ would have bidirectional edges to all nodes in $\sigma$, contradicting the fact that $\sigma$ is a maximal clique.

At the other extreme, consider an oriented graph $G$.  Since $G$ has no bidirectional edges, it contains no cliques of size greater than 1.  Thus the only cliques in $G$ are the singletons, and a singleton is target-free precisely when it has no outgoing edges, i.e. it is a \emph{sink}.  Thus, in the case of oriented graphs, the conjecture reduces to the following theorem from \cite{CTLN-preprint}. 

\begin{theorem}[\cite{CTLN-preprint}]\label{thm:oriented}
Let $G$ be an oriented graph.  Then for any CTLN with graph $G$, the supports of the stable fixed points are precisely the singletons that are sinks in $G$.
\end{theorem}

\begin{remark} A key ingredient to the proof of Theorem~\ref{thm:oriented} in \cite{CTLN-preprint} was the requirement that $\varepsilon < \frac{\delta}{\delta+1}$, which defines the legal parameter range of CTLNs.  Without this condition, there may be nontrivial subgraphs of oriented graphs that can support stable fixed points.  For example, it is straightforward to check that for the $3$-cycle (the oriented graph in Figure~\ref{fig:network-setup}C), the fixed point with support $123$ is stable whenever $\varepsilon > \delta$. 
\end{remark}

These extreme cases provide evidence for Conjecture~\ref{conjecture}, but we are still left with the question of whether the conjecture holds more generally.  Are there other types of graphs where we can guarantee that target-free cliques are the \emph{only} subgraphs that support stable fixed points?  Can we prove that certain types of subgraphs \emph{never} support stable fixed points, irrespective of what graph they are embedded in?  These questions are the primary focus of this paper.  In the following section, we summarize the main results we have proven in this vein, which provide further evidence in support of Conjecture~\ref{conjecture}.

\subsection{Summary of main results}
To state the results, we first set some notation.  We assume throughout the remainder of the paper that we are always considering a nondegenerate CTLN $W=W(G, \varepsilon, \delta)$ with arbitrary parameters within the legal range (unless otherwise noted).  Thus the results are parameter-independent unless explicitly noted otherwise.  The graph $G$ is assumed to have $n$ nodes, and $\sigma \subseteq [n]$. Finally, as shorthand, we often refer to $\sigma$ as having a particular graphical property, when technically we mean that the subgraph $G|_\sigma$ has that property.  

Our first major result shows that among small subgraphs, the only ones that can support stable fixed points are target-free cliques.  Consequently if there are counterexamples to the conjecture, they can only occur in subgraphs of size at least 5.  

\begin{theorem}\label{thm:n4}
If $|\sigma|\leq 4$, then 
$\sigma$ supports a stable fixed point of $W$ if and only if $\sigma$ is a target-free clique in $G$.
\end{theorem}


The proof of Theorem~\ref{thm:n4} is postponed to Section~\ref{sec:n4}, as it relies on the application of many of the other main results to rule out families of subgraphs as possible supports of stable fixed points.  One such collection of graphs that we can rule out are those that are very far from having clique-like structure by way of being relatively sparse.

\begin{theorem}\label{thm:sparse-G}
If $G|_\sigma$ has maximum in-degree $d^{in}_{\max}  < \dfrac{|\sigma|}{2}$, then $\sigma$ cannot support a stable fixed point.
\end{theorem}

At the other extreme, Theorem~\ref{thm:near-cliques} shows that various dense graphs that are ``near-cliques" also cannot support stable fixed points.  Recall that \emph{directed cliques} are graphs that generalize clique structure in directed graphs (see the graph theory terminology in Section~\ref{sec:CTLN-setup} for the precise definition).

\begin{theorem}\label{thm:near-cliques}
Let $\sigma$ be a directed clique or contain a clique of size $|\sigma| -1$.  Then $\sigma$ cannot support a stable fixed point unless $\sigma$ is a clique.
\end{theorem}

Theorem~\ref{thm:sparse-G} is proven in Section~\ref{sec:degree-bounds}, while the proof of Theorem~\ref{thm:near-cliques} is given in Section~\ref{sec:near-cliques}.  In Section~\ref{sec:composite-graphs}, we examine infinite families of graphs, known as \emph{composite graphs}, that are built from connecting arbitrary component subgraphs in a prescribed manner.  We show that when the component subgraphs have certain properties, e.g., sparsity, then the full graph is guaranteed to never support a stable fixed point.  On the flip side, when the component subgraphs are dense, but there are relatively few edges between components, then again the full graph is guaranteed never to support a stable fixed point.  

The results presented so far show that the conjecture holds for certain families of graphs, irrespective of the CTLN parameters. In contrast, the following theorem shows that if we restrict to sufficiently strong inhibition, i.e. sufficiently large $\delta$, then the conjecture holds for all graphs, provided the CTLN lies within this parameter regime.

\begin{theorem}\label{thm:degree-bounds-simple}
Let $G$ be an arbitrary graph and $W=W(G, \varepsilon, \delta)$ be a corresponding CTLN whose parameters satisfy
$$\delta > \frac{\varepsilon}{1-\varepsilon}(n^2-n-1).$$ Then for any $\sigma \subseteq [n]$, $\sigma$ supports a stable fixed point of $W$ if and only if $\sigma$ is a target-free clique in $G$.
\end{theorem}

\noindent The condition on $\delta$ in Theorem~\ref{thm:degree-bounds-simple} is rather extreme, since it grows with $n^2$; Theorem~\ref{thm:degree-bounds} in Section~\ref{sec:degree-bounds} provides more generous bounds on $\delta$ when the size of $\sigma$ or the maximum in-degree of $G|_\sigma$ can be taken into account.  In fact, Theorem~\ref{thm:sparse-G}, which holds for the full legal parameter range, is a corollary of this result.  In each of these cases, though, the requirement of large $\delta$ is an artifact of the techniques used to prove the results; we do not believe that this restriction is actually necessary for the conjecture to hold.  Moreover, we present these results as proof of concept that there exists a parameter regime where the conjecture holds for all graphs, not only for restricted families of graphs with special properties.

\subsection{Bounds on the number of stable fixed points}
In previous work, it was shown that the maximum number of stable fixed points of a CTLN is at most $2^{n-1}$ (Corollary 1 in \cite{fp-paper}).  This bound was derived from constraints on the vector field that follow from the Poincar\'e-Hopf theorem; it does not exploit any graphical structure that might constrain how the subgraphs supporting stable fixed points can coexist within a graph.  Here we show that there is a tighter bound on the number of target-free cliques that can coexist in a graph on $n$ nodes.  Thus, if the conjecture is true, or in any context where it holds, this provides a tighter upper bound on the number of stable fixed points of a CTLN.  

To derive this bound, we make use of a straightforward correspondence between target-free cliques of directed graphs and maximal cliques of undirected graphs (see Section~\ref{sec:maximal-cliques} for more details). This enables us to exploit results on maximal cliques, which have been significantly studied in extremal graph theory.  Applying a result from Moon and Moser \cite{moon-moser}, we immediately obtain the following upper bound on the number of target-free cliques in a directed graph.

\begin{theorem}\label{thm:maximal-cliques}
The maximum number of target-free cliques in a directed graph of size $n$ is
$$ \text{max } \# \text{ of target-free cliques} = \left\{\begin{array}{cl}3^{n/3} & \textrm{if } n \equiv 0 \pmod 3 \\ 
4\cdot 3^{\lfloor n/3 \rfloor -1} &  \textrm{if } n \equiv 1 \pmod 3 \\
2\cdot 3^{\lfloor n/3 \rfloor} &  \textrm{if } n \equiv 2 \pmod 3 .
\end{array}\right.\quad \quad $$
\end{theorem}

\begin{figure}[!bh]
\begin{center}
\includegraphics[width=4.1in]{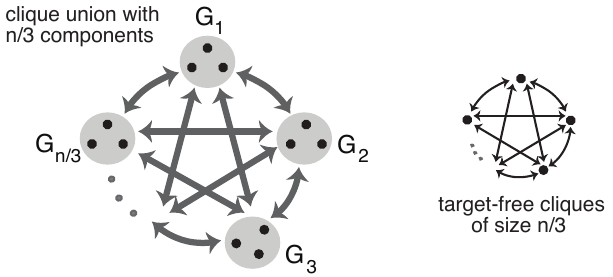}
\vspace{.05in}
\caption{{\bf Graph with maximum number of target-free cliques.} (Left) Cartoon of graph on $n$ nodes, where $n\equiv 0 \pmod 3$, that is a \emph{clique union} of component subgraphs $G_1, \ldots, G_{n/3}$, each of which is an independent set on 3 nodes.  Thick edges indicate that every node in one component send edges to every node in the other component, so there are all-to-all connections between all nodes in different components. (Right) One of the $3^{n/3}$ maximal cliques from the graph on the left.  Each such clique is also target-free.
}
\label{fig:tf-cliques}
\end{center}
\vspace{-.2in}
\end{figure}

Moreover, it is straightforward to construct the graph of a CTLN that attains this upper bound.  Figure~\ref{fig:tf-cliques} illustrates the construction in the case when $n\equiv 0 \pmod 3$.  The graph is comprised of $n/3$ component subgraphs, each containing 3 nodes with no edges between them (an \emph{independent set}).  The components are connected following a \emph{clique union}\footnote{Clique unions are a special type of \emph{composite graph}, first introduced in \cite{fp-paper}, and further considered here in Section~\ref{sec:composite-graphs}.} architecture, in which there is a bidirectional edge between every pair of nodes in different components.   This graph contains $3^{n/3}$ maximal cliques of size $n/3$, each consisting of one node per component.  To see that each of these cliques is target-free, let $\sigma$ be such a clique and consider any $k \notin \sigma$.  
The node $k$ must be in some $G_i$, and since $\sigma$ contains a node from every component subgraph, it must contain a $j$ in $G_i$.  Since $G_i$ is an independent set, $j \not\to k$, and so $k$ cannot be a target of $\sigma$.  Thus, $\sigma$ is target-free.   To modify this construction for the case when $n \equiv 1 \pmod 3$, we simply make one of the component subgraphs contain 4 nodes instead of 3.  For the case when $n \equiv 2 \pmod 3$, we add an additional component subgraph that is an independent set on 2 nodes.

\subsection{Discussion}

A target-free clique $\sigma$ is always minimal in $\FP(W)$ with respect to inclusion because any proper subset of $\sigma$ is necessarily a clique with a target, and thus cannot be in $\FP(W)$ (see Theorem~\ref{thm:cliques-evals}).  Moreover, target-free cliques are a special case of what are known as core motifs,\footnote{Note that $\sigma$ is a \emph{core motif} if it is the unique fixed point of its restricted subnetwork, i.e. $\FP(W_\sigma)=\{\sigma\}$. For any $\tau \subsetneq \sigma$, we see that $\tau \notin \FP(W_\sigma)$ and consequently $\tau \notin \FP(W)$ (see \cite[Corollary 2]{fp-paper} for more details).  Thus, core motifs are always minimal supports within $\FP(W)$. } which have been observed computationally to have a close correspondence with the attractors of CTLNs \cite{core-motifs}.  If Conjecture~\ref{conjecture} is true, then all stable fixed points are guaranteed to be core motifs, and thus minimal in $\FP(W)$.  Even if the conjecture is not true, it may still be the case that all stable fixed points of CTLNs are core motifs or at least minimal.  This may be considered a weaker version of Conjecture~\ref{conjecture}.

It is already known that Conjecture~\ref{conjecture} only applies to CTLNs, as there are TLNs that violate the conjecture in both directions: there are TLNs where (1) a target-free clique of the corresponding graph does not support a stable fixed point, and others where (2) there are stable fixed points supported on other types of subgraphs.  However, it is still an open question whether stable fixed points always correspond to core motifs.  In particular, the following questions remain.\\

\noindent {\bf Question.} For a general TLN $(W,b)$, if $\sigma \in \FP(W,b)$ supports a stable fixed point, is $\sigma$ guaranteed to be a core motif?  If not, is $\sigma$ at least guaranteed to be minimal in $\FP(W,b)$?\\

\noindent We conjecture that the answer to this question is yes, stable fixed points are minimal fixed points.  However, we do not currently have strong evidence in support of this outside of the case of CTLNs, and so it requires further exploration.

\medskip

The remainder of the paper is organized as follows.  Section~\ref{sec:preliminaries} reviews general background on fixed points of CTLNs and then lays out the key techniques to prove a subset cannot support a stable fixed point, which will be used throughout the following sections to prove the main results.  Section~\ref{sec:degree-bounds} is devoted to proving Theorem~\ref{thm:sparse-G}, guaranteeing that sparse graphs never support stable fixed points, and Theorem~\ref{thm:degree-bounds-simple}, showing the conjecture holds within a subset of the legal parameter range.  We will see that both of these results are in fact immediate corollaries of a single technical theorem that connects $|\sigma|$ and the maximum in-degree of $G|_\sigma$ to regions of parameter space in which a fixed point supported on $\sigma$ is guaranteed to be unstable.  Section~\ref{sec:near-cliques} collects various results about ``near-cliques" and concludes with the proof of Theorem~\ref{thm:near-cliques}.  In Section~\ref{sec:composite-graphs}, we consider a broad class of \emph{composite graphs}, which include clique unions as a special case.  We prove that sufficiently sparse composite graphs never support a stable fixed point.  Next, Section~\ref{sec:n4} considers all possible fixed point supports up to size 4 and shows that, among subgraphs that can support fixed points, cliques are the only ones with corresponding fixed points that are stable, proving Theorem~\ref{thm:n4}.  Finally, in Section~\ref{sec:maximal-cliques}, we develop the correspondence between target-free cliques of directed graphs and maximal cliques of undirected graphs, and provide a summary of the literature on the latter to obtain counts of target-free cliques and algorithms for enumerating them.


\section{Preliminaries}\label{sec:preliminaries}
\subsection{General background on fixed points and stability}\label{sec:general-background}
Throughout this work, we will be focused on the subsets $\sigma$ that can support fixed points of a CTLN $W=W(G, \varepsilon, \delta)$, i.e. subsets $\sigma \in \FP(W)$ where
$$\FP(W) \od \{\sigma \subseteq [n] \mid \sigma \text{ is the support of a fixed point}\}.$$ 
Note that we omit $\theta$ from the notation $\FP(W)$ because the value of $\theta >0$ has no impact on the fixed point supports, it simply scales the precise value of the corresponding fixed points. To exploit previous characterizations of fixed points in terms of their supports \cite{fp-paper}, we will restrict consideration to CTLNs that are \underline{nondegenerate}, as defined below.

\begin{definition} \label{def:nondegenerate}
We say that a CTLN $(W,\theta)$ is {\it nondegenerate} if 
\begin{itemize}
\item $\det(I-W_\sigma) \neq 0$ for each $\sigma \subseteq [n]$, and 
\item for each $\sigma \subseteq [n]$ and all $i \in \sigma$,  the corresponding Cramer's determinant is nonzero: $\det((I-W_\sigma)_i;\theta) \neq 0$. 
\end{itemize}
\end{definition}
\noindent Note that almost all CTLNs are nondegenerate, since having a zero determinant is a highly fine-tuned condition.  The notation $\det(A_i; b)$ denotes the determinant obtained by replacing the $i^{\text{th}}$ column of $A$ with the vector $b$, as in Cramer's rule.   In the case of a restricted matrix, $((A_\sigma)_i;b_\sigma)$ denotes the matrix obtained from $A_\sigma$ by replacing the column corresponding to the index $i \in \sigma$ with $b_\sigma$ (note that this is not typically the $i^{\text{th}}$ column of $A_\sigma$).

In \cite{fp-paper}, multiple characterizations of $\FP(W)$ were developed for nondegenerate threshold-linear networks in general as well as CTLNs specifically, including a variety of \emph{graph rules} for CTLNs.  As an immediate consequence of one of these characterizations, it was shown that $\sigma$ is the support of a fixed point, i.e.\ $\sigma \in \FP(W)$, precisely when
\begin{itemize}
\item[(1)] $\sigma \in \FP(W_\sigma)$, and 
\item[(2)]  $\sigma \in \FP(W_{\sigma \cup \{k\}})$ for every $k \notin \sigma$
\end{itemize}
(see \cite[Corollary 2]{fp-paper} for more details).  We say that $\sigma$ is a \emph{permitted motif} of $W$ when it is a fixed point of its restricted subnetwork, so that condition (1) holds.  This condition is satisfied whenever the fixed point corresponding to $\sigma$ has strictly positive entries within $\sigma$, i.e. when $(I-W_\sigma)^{-1}1_\sigma > 0$ (see Equation~\eqref{eq:x_sigma^*}).  

Given a permitted motif $\sigma$, we say that $\sigma$ \emph{survives} to support a fixed point in the full network when condition (2) is satisfied.  Note that whether a subset $\sigma$ is permitted depends only on the subgraph $G|_\sigma$ (and potentially the choice of parameters $\varepsilon$ and $\delta$), while its survival will depend on the embedding of this subgraph in the full graph.  

We say that $\sigma$ is a \emph{stable motif} if it is a permitted motif and its corresponding fixed point is stable, which occurs when all the eigenvalues of $-I+W_\sigma$ have negative real part (or equivalently all the eigenvalues of $I-W_\sigma$ have positive real part) by Theorem~\ref{thm:stable-TLN}.  Thus, stable motifs are the only candidate subsets for supporting stable fixed points, but whether they actually survive to yield stable fixed points of the full network will depend on their embedding.  

\begin{definition}
Let $(W,\theta)$ be a CTLN on $n$ neurons, and let $\sigma \subseteq [n]$. We say that $\sigma$ is a {\it stable motif} of the network if the following two conditions hold:
\begin{itemize}
\item[(i)] $(I-W_\sigma)^{-1}1_\sigma > 0$, and
\item[(ii)] $-I + W_\sigma$ is a stable matrix (i.e., all eigenvalues have negative real part).
\end{itemize}
\end{definition}

\noindent Again, note that $\sigma$ being a stable motif is a necessary, but not sufficient, condition for the existence of a stable fixed point with support $\sigma$ in the full network $(W,\theta)$; one must also check that this stable fixed point actualy \emph{survives} in the full network.

\subsection{Uniform in-degree and simply-added splits} \label{sec:uniform-in-degree}

As our first example of graphs that are guaranteed to be permitted motifs, and thus candidate stable motifs, we turn to those with \emph{uniform in-degree.}

\begin{definition}
We say that $\sigma$ has {\it uniform in-degree} $d$ if every node in $\sigma$ has $d$ incoming edges within $G|_\sigma$, i.e. if the in-degree $d_i^{\mathrm{in}} = d$ for all $i \in \sigma$.     
\end{definition}

\begin{theorem}[Theorem 5 (uniform in-degree) in \cite{fp-paper}] \label{thm:uniform-in-degree}
Suppose $\sigma$ has uniform in-degree $d$ in a graph $G$. Then $\sigma$ is a permitted motif of any CTLN $W$ with graph $G$.\\
For $k \notin \sigma$, let $d_k \od |\{i \in \sigma \mid i \to k\}|$ be the number of edges $k$ receives from $\sigma$.
 Then 
 $$\sigma \in \FP(W) \;\; \Leftrightarrow \;\; d_k \leq d \text{ for all } k \notin \sigma.$$ 
Furthermore, if $d <|\sigma|/2$ and $|\sigma|>1$, then the fixed point is unstable.  If $d=|\sigma|-1$, i.e.\ if $\sigma$ is a clique, then the fixed point is stable.
\end{theorem}

The particular proof techniques used for Theorem~\ref{thm:uniform-in-degree} only enabled us to prove that $\sigma$ is unstable when $d < |\sigma|/2$, but we conjecture that this holds whenever $d<|\sigma| -1$, i.e.\ whenever $\sigma$ is not a clique.

Combining Theorem~\ref{thm:uniform-in-degree} with a straightforward computation of eigenvalues yields the following result characterizing cliques.

\begin{theorem}[\cite{fp-paper}]\label{thm:cliques-evals}
Let $\sigma$ be a clique in $G$ and let $W$ be a CTLN with graph $G$.  Then 
 $$\sigma \in \FP(W) \;\; \Leftrightarrow \;\; \sigma \text{ is target-free.}$$ 
Moreover, $\sigma$ is a stable motif, with eigenvalues of $I-W_\sigma$ given by
$$\eig(I-W_\sigma) = \{\varepsilon,~|\sigma|(1-\varepsilon)-\varepsilon\},$$
where $\varepsilon$ has multiplicity $|\sigma| -1$.  
\end{theorem}

One graph structure that will prove particularly useful for getting a handle on eigenvalues of larger motifs is that of \emph{simply-added splits}.  In particular, if a motif has a simply-added split containing a uniform in-degree subgraph, then the eigenvalues of the uniform in-degree subgraph will be inherited as eigenvalues of the full motif (Lemma~\ref{lemma:simply-added-evals}).

\begin{figure}[!h]
\begin{center}
\includegraphics[height=1.5in]{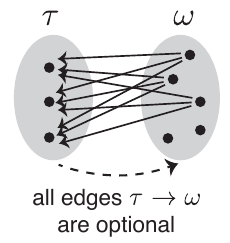}
\vspace{-.15in}
\end{center}
\caption{In this graph, $\omega$ is simply-added to $\tau$ and thus each $i \in \omega$ either sends all possible edges to $\tau$, or no edges. There are no constraints on the edges within $\tau$, within $\omega$, or from $\tau$ to $\omega$.}
\label{fig:simply-added}
\end{figure}

Suppose $\sigma = \tau\;  \dot\cup \;\omega$ with $\tau,\; \omega \neq \emptyset$ and $\tau \cap \omega = \emptyset$.  We say that $\omega$ is \emph{simply-added} onto $\tau$ (or equivalently that $\sigma$ has a \emph{simply-added split} $\tau\;  \dot\cup \;\omega$) if each node in $\omega$ treats all the nodes in $\tau$ identically in terms of its outgoing edges, i.e.\ for each $i \in \omega$ if $i \to j$ for some $j \in \tau$, then $i \to j$ for \underline{every} $j \in \tau$ (see Figure~\ref{fig:simply-added}).  Note that there are no constraints on the edges from $\tau$ back to $\omega$ or on edges within $\tau$ or $\omega$.

\begin{lemma}[Lemma 5 in \cite{fp-paper}]\label{lemma:simply-added-evals}
Suppose $\sigma$ has a simply-added split $\tau\;  \dot\cup \;\omega$ with $\omega$ simply-added to $\tau$, where $\tau$ is uniform in-degree. Let $R_\tau$ be the row sum of $I-W_\tau$. Note that this is the maximum (Perron-Frobenius) eigenvalue of $I-W_\tau$. Then 
$$\eig(I-W_\sigma) \supset \eig(I-W_\tau) \setminus R_\tau.$$ 
So all the eigenvalues of $\tau$ get inherited, except possibly the top one $R_\tau$. 
\end{lemma}

As a consequence of Lemma~\ref{lemma:simply-added-evals} and \cite[Theorem 1.4]{sequential-att-paper}, we obtain the following corollary that rules out certain graphs as stable motifs.  

\begin{corollary}\label{cor:simply-added-unstable}
Suppose $\sigma$ has a simply-added split $\tau\;  \dot\cup \;\omega$ with $\omega$ simply-added to $\tau$, where $\tau$ is uniform in-degree. If $\tau$ is not a stable motif, then $\sigma$ is not a stable motif.   
\end{corollary}


\subsection{Techniques for ruling out stable motifs}\label{sec:ruling-out-stable}
In this section, we highlight and synthesize background results that underlie our arguments ruling out graphs as candidate stable motifs.  Corollary~\ref{cor:simply-added-unstable} provides one such result in this direction, whenever the graph has a simply-added split with an unstable uniform in-degree subset.  More generally, to rule out a $\sigma$ as a stable motif, we can either show that $\sigma$ is not permitted or show that $-I+W_\sigma$ has an eigenvalue with positive real part (equivalently, $I-W_\sigma$ has an eigenvalue with negative real part).  \emph{Graphical domination}, first defined in \cite{fp-paper}, is the primary method for showing that $\sigma$ is not a permitted motif.

\begin{definition}[\cite{fp-paper}]\label{def:graph-domination}
Let $G$ be a graph on $n$ nodes and $\sigma \subseteq [n]$.  For $j, k \in \sigma$, we say that $k$ \emph{graphically dominates} $j$ with respect to $\sigma$ if the following three conditions all hold:
$$(1) \ \  \text{if } i \to j \text{ then } i \to k, \text{ for every } i \in \sigma \setminus \{j, k\}, \quad  \quad  (2)  \ \  j \to k,  \quad  \quad (3)  \ \  k \not\to j \quad \quad $$
\end{definition}

\begin{theorem}[Theorem 4 (graphical domination) in \cite{fp-paper}]\label{thm:graph-domination}
Let $G$ be a graph on $n$ nodes and $\sigma \subseteq [n]$. Suppose $k$ graphically dominates $j$ with respect to $\sigma$ for some $j, k \in \sigma$.  Then $\sigma$ is not a permitted motif, and so $\sigma \notin \FP(W)$ for any CTLN $W$ with graph $G$. 
\end{theorem}

Another method for ruling out $\sigma$ as a stable motif is \emph{parity} (Theorem~\ref{thm:parity} below).  Throughout the following, to simplify analyses, we will focus on the matrix $I-W_\sigma$; then $\sigma$ is stable precisely when all the eigenvalues of $I-W_\sigma$ have positive real part.  We define the \emph{index} of $\sigma$ as 
$$\idx(\sigma) \od \sgn \det (I-W_\sigma).$$
Since the determinant is the product of the eigenvalues, we see that any stable motif $\sigma$ must have $\idx(\sigma)=+1$.  Thus, any $\sigma$ with $\idx(\sigma)=-1$ is guaranteed to not be a stable motif.

The following theorem, which is a direct consequence of the Poincar\'e-Hopf Theorem, shows that there is a detailed balance between the indices of all the fixed points of a CTLN.

\begin{theorem}[parity \cite{CTLN-paper}]\label{thm:parity}
Let $W=W(G, \varepsilon, \delta)$ be a CTLN. Then
$$\sum_{\sigma \in \FP(W)} \idx(\sigma) = + 1.$$
In particular, the total number of fixed points $|\FP(W)|$ is always odd.
\end{theorem}

As an immediate consequence of Theorem~\ref{thm:parity}, we see that a CTLN has at most $2^{n-1}$ stable fixed points, since these all have index $+1$ and the maximum size of $\FP(W)$ is $2^n-1$, the number of choices for a nonempty support $\sigma$.  

Beyond this general upper bound, Theorem~\ref{thm:parity} can be used to explicitly rule out graphs as stable motifs whenever we know which proper subgraphs are permitted motifs that survive to yield fixed points of the full graph.  Specifically, if we know that there are an odd number of proper subgraphs of a graph $G$ that survive as fixed points, then by Theorem~\ref{thm:parity}, $G$ cannot have a full-support fixed point and thus $G$ is not a permitted motif.  If we additionally know the index of all the subgraphs of a graph $G$ that survive as fixed points, the following conclusions can immediately be drawn from the sum of these indices:
\[
\displaystyle \sum_{\sigma \in \FP(W);~\sigma \neq [n]} \hspace{-.2in} \idx(\sigma) = \left\{ \begin{array}{rcl}+1 &\quad  \Rightarrow & G \text{ is not a permitted motif}\\
 +2 &\quad  \Rightarrow & G \text{ is permitted, but not stable}\\
0 &\quad  \Rightarrow & G \text{ is a permitted motif and stability is unknown}\\
\end{array}\right.
\]

The following lemma from \cite{fp-paper} gives another tool for determining a motif's index from that of its proper subgraphs of size one less.  In particular, this lemma shows that a CTLN can never contain two stable fixed points whose support differ in only a single neuron. 

\begin{lemma}[Lemma 3 (alternation) in \cite{fp-paper}]\label{lemma:alternation}
Let $W=W(G, \varepsilon, \delta)$ be a CTLN.  If $\sigma, \sigma \cup \{k\} \in \FP(W),$ for $k \notin \sigma,$ are both fixed point supports, then 
$$\idx(\sigma\cup\{k\}) = - \idx(\sigma).$$
\end{lemma}

For the previous set of tools, we exploited the fact that the determinant of a matrix equals the product of its eigenvalues.  For the remainder of this section, we focus on tools that will utilize the fact that the trace of a matrix is the sum of its eigenvalues.  We begin with the following key observation we will exploit throughout.  \\

\noindent\textbf{Observation.} If the sum of a subset of eigenvalues of $I-W_\sigma$ is bigger than its trace, $\Tr(I-W_\sigma)=|\sigma|$, then $I-W_\sigma$ must have a negative eigenvalue.  In this case, the matrix $-I+W_\sigma$ is unstable, and so $\sigma$ is not a stable motif.  In particular, if the maximum eigenvalue $\lambda_{\max}$ of $I-W_\sigma$ is larger than $|\sigma|$, then $\sigma$ is not a stable motif.  \\

We next collect some useful results for lower bounding the maximum eigenvalue $\lambda_{\max}$.

\begin{theorem}[Collatz-Wielandt]\label{thm:C-W}
Let $A$ be an $n \times n$ matrix with strictly positive entries.  Then the Perron-Frobenius eigenvalue $\lambda_{\max}$ of $A$ is given by
$$\lambda_{\max} = \displaystyle \max_{x \in \RR^n_{\geq 0}} \ \left(\min_{i \in [n]} \frac{[Ax]_{i}}{x_{i}} \right),$$
where the maximum is taken over all non-negative vectors $x$.  
In particular, $$\lambda_{\max} \geq \min_{i \in [n]} \frac{[Ax]_{i}}{x_{i}},$$ for any non-negative vector $x$.  
\end{theorem}

\begin{lemma}\label{lemma:lambda_max}
Let $A$ and $B$ be $n \times n$ matrices with strictly positive entries, and suppose $A \geq B$ entrywise.  Then
$$\lambda_{\max}(A) \geq \lambda_{\max}(B).$$
\end{lemma}
\begin{proof}
Since $A \geq B$, we have $\frac{[Ax]_{i}}{x_{i}} \geq \frac{[Bx]_{i}}{x_{i}}$ for every non-negative vector $x$.  Thus, by Theorem~\ref{thm:C-W} (Collatz-Wielandt), $\lambda_{\max}(A) \geq \lambda_{\max}(B)$. 
\end{proof}


Since entrywise comparison of CTLN matrices is fully determined by edges in the underlying graphs, we immediately obtain the following corollary.  Note that $E(G)$ denotes the set of edges of a graph $G$.

\begin{corollary}
Suppose that $G$ and $G'$ are two graphs on $n$ nodes satisfying $E(G) \subseteq E(G')$, i.e.\ $i \to j$ in $G$ implies $i \to j$ in $G'$.  Let $W$ and $W'$ be CTLNs for graphs $G$ and $G'$ respectively.  Then
$$\lambda_{\max}(I-W) \geq \lambda_{\max}(I-W').$$
\end{corollary}
\begin{proof}
Observe that the off-diagonal entries of $W$ and $W'$ are all $1-\varepsilon$ and $1+\delta$.  Since $E(G) \subseteq E(G')$, whenever $(I-W)_{ij}$ takes on the smaller value $1-\varepsilon$ due to $j \to i$ in $G$, we have $(I-W')_{ij}$ equals $1-\varepsilon$ as well.  Thus, $I-W \geq I-W'$ entrywise, and so by Lemma~\ref{lemma:lambda_max} the result follows.  
\end{proof}


\section{Degree bounds and proofs of Theorems~\ref{thm:sparse-G} and~\ref{thm:degree-bounds-simple}}\label{sec:degree-bounds}
In this section, we prove Theorem~\ref{thm:sparse-G} showing that sparse graphs cannot be stable motifs as well as Theorem~\ref{thm:degree-bounds-simple}, which guarantees that there is a parameter regime in which the only stable motifs are cliques, and thus the conjecture holds within this regime.  Interestingly, these results are both corollaries of a single theorem that connects properties of $\sigma$, such as its size or maximum in-degree, to a range of $\delta$ values in which $\sigma$ can never be a stable motif. The key to the proof of this theorem is to show that whenever $\delta$ lies in the relevant range, $I-W_\sigma$ has an eigenvalue that is larger than the trace, forcing a negative eigenvalue.  Toward that end, we begin with a formula for the Perron-Frobenius eigenvalue $\lambda_{\max}$ of $I-W_\sigma$, when $\sigma$ is exactly one edge away from a clique.   To compute the eigenvalue of such a CTLN, we begin with a lemma identifying an eigenvalue and eigenvector of a highly structured matrix.  The proof of this lemma is a straightforward computation, and thus is left to the reader as an exercise.

\begin{lemma}\label{lemma:eigenvalue}
Let $A$ be an $n \times n$ matrix of the form 
$$A= \begin{bmatrix}
1 & a & \ldots& a & b \\
a & 1 & \ldots & a & a \\
\vdots  & & \ddots & \vdots & \vdots\\
a & a  & \ldots & a& 1
\end{bmatrix}
$$
for any $a, b \in \RR$.    Then $A$ has an eigenvector $[c~1~1~\ldots~1]^t$ with corresponding eigenvalue $\lambda$, where
$$
c = 1 + \frac12\left[ -n + \sqrt{n^2 + 4 \left(\frac{b-a}{a}\right)}\right], \quad \lambda= 1-a + \frac{a}{2}\left[n + \sqrt{n^2 + 4 \left(\frac{b-a}{a}\right)}\right] 
$$
Furthermore, if $b \geq a >0$, then $[c~1~1~\ldots~1]^t$ is the Perron-Frobenius eigenvector and $\lambda=\lambda_{\max}$.  
\end{lemma}

As a special case of Lemma~\ref{lemma:eigenvalue}, we obtain the formula for an eigenvalue of a graph that is a clique missing exactly one edge.  

\begin{corollary}\label{corollary:clique-minus-edge}
Let $\sigma$ be a clique missing exactly one edge, $j \not\to i$.  Then $I-W_\sigma$ has Perron-Frobenius eigenvalue
$$\lambda_{\max}= \varepsilon+\frac{1}{2}(1-\varepsilon)(|\sigma|+\sqrt{|\sigma|^2+4(\delta+\varepsilon)/(1-\varepsilon)}).$$
\end{corollary}
\begin{proof}
Since $\sigma$ is a clique missing exactly one edge, without loss of generality suppose the missing edge is from the last node to the first node in $\sigma$, i.e.\ $|\sigma| \not\to 1$.  Then $I-W_\sigma$ has the form of the matrix $A$ in Lemma~\ref{lemma:eigenvalue} with $a=1-\varepsilon$ and $b=1+\delta$.  Since $b \geq a>0$, the value of $\lambda_{\max}$ follows directly from that result.  
\end{proof}

\smallskip

The following theorem gives parameter regimes in which non-cliques are guaranteed to never be stable motifs.

\begin{theorem}\label{thm:degree-bounds}
Let $G$ be an arbitrary graph and $W=W(G,\varepsilon, \delta)$ a corresponding CTLN.  For $\sigma \subseteq [n]$, if $\sigma$ is not a clique, then $\sigma$ is not a stable motif whenever the CTLN parameters satisfy
$$\delta > \frac{\varepsilon}{1-\varepsilon}(|\sigma|^2-|\sigma|-1).$$
Moreover, if $G|_\sigma$ has maximum in-degree $d^{in}_{\max} =d < |\sigma|-1$, then $\sigma$ is not a stable motif whenever
$$\delta >  \frac{\varepsilon d}{|\sigma|-1-d}.$$
\end{theorem}
\begin{proof}
Let $\sigma$ be any subset that is not a clique.  Let $G'$ be a graph of size $|\sigma|$ that is a clique missing exactly one edge, as in Corollary~\ref{corollary:clique-minus-edge}.  Since $\sigma$ is not a clique, $E(G|_\sigma) \subseteq E(G')$, and so by Lemma~\ref{lemma:lambda_max}, 
$$\lambda_{\max}(I-W_\sigma) \geq \lambda_{\max}(I-W') =\varepsilon+\frac{1}{2}(1-\varepsilon)(|\sigma|+\sqrt{|\sigma|^2+4(\delta+\varepsilon)/(1-\varepsilon)}).$$
Recall that whenever
$$\lambda_{\max}(I-W_\sigma)>|\sigma| = \Tr(I-W_\sigma), $$
$I-W_\sigma$ must have a negative eigenvalue, forcing $-I+W_\sigma$ to be unstable.  
Solving for $\delta$, it is straightforward to show that $\varepsilon+\frac{1}{2}(1-\varepsilon)(|\sigma|+\sqrt{|\sigma|^2+4(\delta+\varepsilon)/(1-\varepsilon)}) > |\sigma|$ whenever 
$$\delta > \frac{\varepsilon(|\sigma|^2 - |\sigma| -1 + (2-|\sigma|)\varepsilon)}{1-\varepsilon}.$$
Since $\sigma$ is not a clique, $|\sigma| \geq 2$, and so it suffices to have $\delta > \frac{\varepsilon }{1-\varepsilon}(|\sigma|^2-|\sigma|-1)$ to guarantee the previous inequality holds.  Then $-I+W_\sigma$ is unstable, and so $\sigma$ is not a stable motif.

To prove the second statement, suppose $\sigma$ has maximum in-degree $d^{in}_{\max} =d < |\sigma|-1$.  Let $G'$ be a graph of size $|\sigma|$ with uniform in-degree $d$ such that $E(G|_\sigma) \subseteq E(G')$.\footnote{Note that it is always possible to find such a $G'$ since there are no constraints on the out-degree, so arbitrary edges can be added to $G|_\sigma$ to appropriately fill up the in-degrees.}  To complete the proof, we follow a similar argument as above to ensure that 
$$\lambda_{\max}(I-W_\sigma) \geq \lambda_{\max}(I-W') >|\sigma| = \Tr(I-W_\sigma), $$
and thereby force $I-W_\sigma$ to have a negative eigenvalue.  In this case, we obtain a tighter bound using the fact that when $G'$ has uniform in-degree $d$, $I-W'$ has the all-ones vector as the Perron-Frobenius eigenvector and thus, $\lambda_{\max}(I-W')$ equals the row sum:
$$\lambda_{\max}(I-W') = |\sigma|+ (|\sigma|-1-d)\delta - d\varepsilon.$$
Solving $\lambda_{\max}(I-W')>|\sigma|$ for $\delta$, we see that whenever
$$\delta >  \frac{\varepsilon d}{|\sigma|-1-d}$$
$I-W_\sigma$ is forced to have a negative eigenvalue, and so $\sigma$ is not a stable motif.
\end{proof}

\smallskip
\noindent From Theorem~\ref{thm:degree-bounds}, we obtain Theorem~\ref{thm:sparse-G} and Theorem~\ref{thm:degree-bounds-simple} as immediate corollaries.

\begin{proof}[{\bf Proof of Theorem~\ref{thm:sparse-G}}]
Let $\sigma \subseteq [n]$ such that $G|_\sigma$ has maximum in-degree $d \od d^{in}_{\max} < \dfrac{|\sigma|}{2}$. Then the bound on $\delta$ in the second part of Theorem~\ref{thm:degree-bounds} reduces to 
$$\frac{\varepsilon d}{|\sigma|-1-d} \leq  \frac{\varepsilon(|\sigma|-1)/2}{|\sigma|-1-(|\sigma|-1)/2} = \varepsilon.$$
Within the legal parameter range, we are guaranteed that $\varepsilon <1$, and $\delta > \frac{\varepsilon}{1- \varepsilon}> \varepsilon$.  Thus, $\delta$ satisfies the bound in Theorem~\ref{thm:degree-bounds} across the full legal range of parameters, and so $\sigma$ is not a stable motif, meaning it cannot support a stable fixed point.
\end{proof}

\begin{proof}[{\bf Theorem~\ref{thm:degree-bounds-simple}}]
When the CTLN parameters satisfy $\delta > \frac{\varepsilon}{1-\varepsilon}(n^2-n-1)$, we are guaranteed that $\delta > \frac{\varepsilon}{1-\varepsilon}(|\sigma|^2-|\sigma|-1)$ for all $\sigma \subseteq [n]$.  Thus, by Theorem~\ref{thm:degree-bounds}, $\sigma$ is not a stable motif whenever $\sigma$ is not a clique.  Thus, the only stable motifs in this parameter regime are the cliques, and these produce stable fixed points if and only if they are target-free (Theorem~\ref{thm:uniform-in-degree}).
\end{proof}

\section{Near-cliques and proof of Theorem~\ref{thm:near-cliques}}\label{sec:near-cliques}

Theorem~\ref{thm:degree-bounds} showed that there is a parameter-regime where non-cliques are provably not stable motifs, and that for sparser graphs we can prove instability for a significantly larger parameter range.  In this section, we consider the other extreme of particularly dense graphs and prove instability of those motifs as well.  Specifically, we build up the machinery to prove Theorem~\ref{thm:near-cliques} showing that directed cliques and graphs containing a clique of size one less are never stable motifs unless they are themselves cliques.   We begin by proving an important fact about the structure of directed cliques.

\begin{lemma}\label{lemma:unique-tf-clique}
Let $\tau \subseteq [n]$ be a directed clique of a graph $G$.  Then there exists a unique $\sigma \subseteq \tau$ such that $\sigma$ is a target-free clique in $G|_\tau$.
\end{lemma}
\begin{proof}
First, we show the existence of a target-free clique in $G|_\tau$.  
Since $\tau$ is a directed clique, we can label the neurons $1,\ldots,t$, where $t=|\tau|$, such that $i \rightarrow j$ whenever $i<j$.
If node $t$ is a sink in $G|_\tau$, then the singleton $\{t\}$ is a target-free clique.  Otherwise $t$ has at least one outgoing edge in $\tau$; let $i_1 \in \tau$ be the vertex with largest index such that $t \rightarrow i_1$, so that $\{i_1, t\}$ is a clique.  Either this is a target-free clique, or it has at least one target in $\tau$.  In the latter case, let $i_2$ be the target of $\{i_1, t\}$ with the largest index.  Clearly, $i_2 < i_1, t$, and so we also have $i_2 \rightarrow i_1$ and $i_2 \rightarrow t$, making $\{i_1,i_2, t\}$ a clique.  If it's not a target-free clique, we can again repeat the previous step and add its target of largest index, yielding a larger clique.  Eventually, we obtain a clique $\sigma = \{i_1,\ldots, i_k, t\}$ that has no targets in $\tau$, and is thus a target-free clique in $G|_\tau$.  (If $\tau$ is itself a clique, then $\sigma = \tau$.)

To see that $\sigma$ is unique, suppose $\omega$ is another target-free clique of $G|_\tau$.  Then $\omega$ must contain the vertex $t$, otherwise $t$ would be a target of $\omega$ by the edge rule defining directed cliques.  Observe that all other elements of $\omega$ must be less than or equal to $i_1$ since all nodes in $\omega$ receive an edge from $t$ and $i_1$ is the largest such element by construction.  Thus $j \to i_1$ for all $j \in \omega$, since $\tau$ is a directed clique, and so $i_1 \in \omega$ since otherwise $\omega$ would have a target.
Similarly, since $\omega$ is a clique containing $\{i_1, t\}$, we must have $i_2 \in \omega$, since otherwise $i_2$ would be a target.  Continuing in this manner we see that $\sigma \subseteq \omega$.   Furthermore, there cannot be a $j \in \omega \setminus \sigma$, since $\omega$ being a clique would force $j$ to be a target of $\sigma$, but $\sigma$ is target-free.  Thus $\sigma = \omega$, and so $\sigma$ is unique.
\end{proof}

\begin{proposition}\label{prop:directed-cliques}
Let G be a directed clique and $W$ be any CTLN with graph G. Then $\FP(W)~=~\{\sigma\}$, where $\sigma$ is the unique target-free clique of G. 
In particular, a directed clique is a stable motif if and only if it is a clique.
\end{proposition}
\begin{proof}
By Lemma~\ref{lemma:unique-tf-clique}, $G$ contains a unique target-free clique $\sigma$, and so $\sigma \in \FP(W)$ and the corresponding fixed point is stable.  To show that $\sigma$ is the only fixed point support, consider any other subset $\omega \subseteq [n]$.   If $\omega$ is a clique, then it necessarily has a target, since $\sigma$ is the unique target-free clique, and thus $\omega \notin \FP(W)$ by Theorem~\ref{thm:cliques-evals}.  Otherwise, $\omega$ is a directed clique that is not a clique, under the same ordering of the vertices that showed $G$ was a directed clique.  Thus, there exists some pair of vertices with only a unidirectional edge between them.  Let $j \in \omega$ be the vertex of largest index such that there exists a $k \in \omega$ with $j \rightarrow k$ but $k \not\rightarrow j$, then choose $k$ to have the largest index among the vertices that do not send edges to $j$ (note by the directed clique property, we necessarily have $k >j$).  We will show that $k$ graphically dominates $j$ with respect to $\omega$.  Consider $i \in \omega$ such that $i \rightarrow j$.  If $i < k$, then $i \rightarrow k$ since $\omega$ is a directed clique.  If $i >k$, we have $k \rightarrow i$ and so we must have $i \rightarrow k$ as well, since otherwise this would contradict the fact that $j$ was the vertex of largest index that has only a unidirectional edge with some node greater than it.  Thus, we have (1) $i \rightarrow j$ implies $i \rightarrow k$  for all $i \in \omega\setminus\{j,k\}$, (2) $j \rightarrow k$, and (3) $k \not\rightarrow j$, and so $k$ graphically dominates $j$ with respect to $\omega$.  Thus, $\omega \notin \FP(W)$ by Theorem~\ref{thm:graph-domination}, and so $\FP(W) = \{\sigma\}$.

Finally, observe that if $G$ is not itself a clique, i.e.\ $[n] \neq \sigma$, then $[n] \notin \FP(W)$, and so $G$ is not a permitted motif and thus cannot be a stable motif.  
\end{proof}

Proposition~\ref{prop:directed-cliques} completes the proof of the first part of Theorem~\ref{thm:near-cliques} by showing that directed cliques are only stable motifs when they are themselves cliques.  To prove the second part of the theorem, we first need the following lemma showing alternation of the index of fixed point supports that differ in one node.

\begin{lemma}\label{lemma:alternation-ufd}
Suppose $\sigma$ is a permitted motif such that $\sigma = \tau \cup \{k\}$ where $\tau$ has uniform in-degree $d$ and the in-degree of $k$ satisfies $d_k^{in} \leq d$.  Then $\idx(\sigma) = - \idx(\tau)$.  
\end{lemma}
\begin{proof}
Let $W$ be any CTLN with graph $G|_\sigma$.  By hypothesis, $\sigma=\tau \cup \{k\}$ is permitted, and so $\tau \cup \{k\} \in \FP(W_\sigma)$.  Additionally $\tau \in \FP(W_\sigma)$ by Theorem~\ref{thm:uniform-in-degree} (uniform in-degree), since $d_k^{in} \leq d$.  Thus by Lemma~\ref{lemma:alternation} (alternation), $\idx(\tau \cup \{k\}) = - \idx(\tau)$.
\end{proof}

Recall that a necessary condition for $\sigma$ to be a stable motif is that $\sigma$ be a permitted motif with index $+1$.   The following proposition shows that if $\sigma$ contains a clique of size 1 less, then it can only be a stable motif if it is itself a clique.  

\begin{proposition}\label{prop:contains-clique}
If $\sigma$ contains a clique of size $|\sigma|-1$ and $\sigma$ is not a clique, then $\sigma$ is not a stable motif.  
\end{proposition}

\begin{proof}
Let $\sigma = \tau \cup \{k\}$ where $\tau$ is a clique of size $|\sigma|-1$, so $\tau$ has uniform in-degree $|\sigma|-2$.  By Lemma~\ref{lemma:alternation-ufd}, if $k$ has in-degree $d_k^{in} \leq |\sigma|-2$, then whenever $\sigma$ is permitted it has index $\idx(\sigma) = - \idx(\tau) = -1$. Thus, when $d_k^{in} \leq |\sigma|-2$, if $\sigma$ is permitted it must be unstable, and so $\sigma$ is not a stable motif.  On the other hand, when $d_k^{in} = |\sigma|-1$, so that $k$ receives from every node in $\tau$, there must exist a $j \in \tau$ such that $k \not\to j$, since $\sigma$ is not a clique.  Then $k$ graphically dominates $j$ with respect to $\sigma$ because condition (1) and (2) are trivially satisfied since $i \to k$ for all $i \in \sigma\setminus\{k\}$ and (3) holds by choice of $j$.  Thus, $\sigma$ is not a permitted motif, and in particular not a stable motif, since it contains a graphical domination relationship.  
\end{proof}

We can now complete the proof of Theorem~\ref{thm:near-cliques}.

\begin{proof}[{\bf Proof of Theorem~\ref{thm:near-cliques}}]
Proposition~\ref{prop:directed-cliques} guarantees that a directed clique is a stable motif if and only if it is a clique, while Proposition~\ref{prop:contains-clique} shows that any graph containing a clique of size 1 less cannot be a stable motif unless it is a clique itself.  
\end{proof}

\section{Composite graphs that are never stable motifs}\label{sec:composite-graphs}

In this section, we prove that a variety of \emph{composite graphs} can never be stable motifs, including some well known constructions such as disjoint unions and cyclic unions.

\begin{figure}[!ht]
\begin{center}
\includegraphics[width=.85\textwidth]{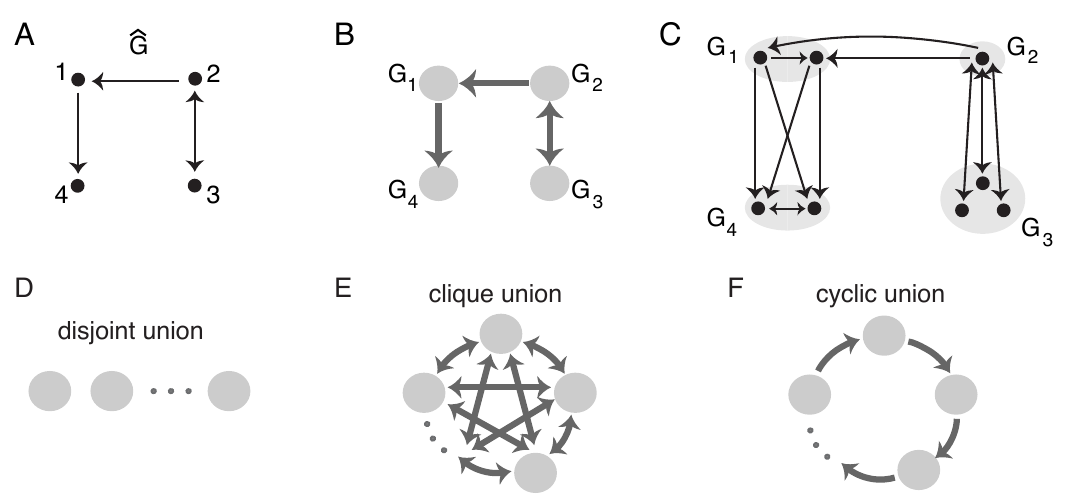}
\caption{(A) A skeleton graph $\widehat G$.  (B) An arbitrary composite graph with skeleton $\widehat G$ from A.  Each node $i$ in the skeleton is replaced with a component graph $G_i$ whose connections to the rest of the graph are prescribed by the connections of node $i$ in $\widehat G$.  (C) An example composite graph with skeleton $\widehat G$ from A. (D-F) Families of composite graphs that have previously  been studied extensively in \cite{fp-paper, sequential-att-paper}.
}
\label{fig:general-composite}
\end{center}
\vspace{-.1in}
\end{figure}

\begin{definition}[composite graph] Given a set of graphs $G_1,\ldots,G_m$, and a graph $\widehat{G}$ on $m$ nodes, the {\it composite graph} with {\it components} $G_i$ and {\it skeleton} $\widehat{G}$ is the graph $G$ constructed by taking the union of all component graphs, and adding edges between components according to the following rule: if $u \in G_i$ and $v \in G_j$, then $u \to v$ in $G$ if and only if $i \to j$ in $\widehat{G}$. (See Figure~\ref{fig:general-composite}.)
\end{definition}

A key feature of composite graphs is that for each component $G_i$, the rest of the graph treats all the nodes in the component identically in terms of the edges projected to that component.  Specifically, in a composite graph, the rest of the graph is simply-added onto each component, so that $G_i \ \dot\cup\ (G\setminus G_i)$ is a \emph{simply-added split} for each component (see Section~\ref{sec:uniform-in-degree}).  This fact was key to proving the following result from \cite{fp-paper} that enables us to immediately rule out certain composite graphs as stable motifs.

\begin{theorem}[Theorem 8 in \cite{fp-paper}]\label{thm:one-bad-apple}
Let $G$ be a composite graph of components $G_{1},\ldots,G_{m}.$ If some component $G_i$ is not a permitted motif, then $G$ is not a permitted motif.  
In particular, $G$ is not a stable motif.  
\end{theorem}

As another consequence of this simply-added split in composite graphs, we can apply Lemma~\ref{lemma:simply-added-evals} whenever one of the components has uniform in-degree.  Recall that $\sigma$ is \emph{uniform in-degree} if all nodes in $G|_\sigma$ have the same in-degree. In this case, all the eigenvalues of the uniform in-degree component are inherited by the full composite graph, except for possibly the Perron-Frobenius eigenvalue $\lambda_{\max}$ of the component.  Thus, if a component is an unstable uniform in-degree, then the full composite graph will also be unstable, as captured by the following proposition from \cite{fp-paper}.  

\begin{proposition}[Proposition 1 in \cite{fp-paper}]\label{prop:composite-unstable}
Let $G$ be a composite graph with components $G_1, \ldots, G_m$.  If some component $G_i$ is an unstable uniform in-degree motif, then $G$ is not a stable motif.
\end{proposition}

These results enable us to rule out as possible stable motifs all composite graphs where either a component is not permitted or it is an unstable uniform in-degree.  But what about the case when all the components are permitted and stable, specifically, what if all the components are cliques?  The following theorem shows that as long as the composite graph satisfies a particular bound on the in-degree of the nodes, it will not be a stable motif even when all the components are stable motif cliques.  

\begin{theorem}\label{thm:composite-unstable}
Let $G$ be a composite graph on $n$ nodes with skeleton $\Ghat$ on $m$ nodes, and suppose the components $G_1, \ldots, G_m$ are all cliques (of arbitrary size).  Let $d_{\max}^{in}(G)$ be the maximum in-degree of all nodes in $G$.  If 
$$d_{\max}^{in}(G) \leq n-\frac{m+1}{2},$$
then $G$ is not a stable motif.  In particular, if the skeleton $\Ghat$ has maximum in-degree $d_{\max}^{in}(\Ghat)~<~\frac{m}{2},$ then $G$ is not a stable motif.
\end{theorem}

\begin{proof}
Let the components $G_1, \ldots, G_m$ be cliques of sizes $n_1, \ldots, n_m$ respectively so that $n = n_1 + \ldots + n_m$.  Recall that a clique $\sigma$ has $\lambda_{\max} = |\sigma|(1-\varepsilon) +\varepsilon$, while the remaining $|\sigma|-1$ eigenvalues all equal $\varepsilon$ (see Theorem~\ref{thm:cliques-evals}).  Since the rest of the graph is simply-added onto each component, Lemma~\ref{lemma:simply-added-evals} guarantees that $I-W$ inherits $n_i -1$ eigenvalues, all equal to $\varepsilon$, from each component $G_i$.  Hence, $\varepsilon$ is an eigenvalue of $I-W$ with multiplicity $\sum_{i=1}^m (n_i -1) = n-m$.  

We will show that the sum of $\lambda_{\max}(I-W)$ with these $n-m$ eigenvalues $\varepsilon$ is strictly greater than $\Tr(I-W) = n$, guaranteeing that $I-W$ has a negative eigenvalue and $G$ is not a stable motif.  Recall that by Theorem~\ref{thm:C-W} (Collatz-Wielandt), 
$\lambda_{\max} \geq \min_{i \in [n]} \frac{[(I-W)x]_i}{x_i},$
 for every nonnegative vector $x$.  We will use this to show that $\lambda_{\max} > n- (n-m)\varepsilon.$  Observe that when $x$ is the all-ones vector, $\frac{[(I-W)x]_i}{x_i}$ is simply the $i^\text{th}$ row sum $R_i$ of $I-W$, which is determined by the number of inputs $d_i^{in}$ to node $i$ in $G$.  Thus, for $x=[1 \ldots 1]^t$, we have
$$\frac{[(I-W)x]_i}{x_i} = R_i = n - \varepsilon d_i^{in} - \delta(n-1-d_i^{in}).$$
Next observe that $R_i > n- (n-m)\varepsilon$ precisely when
\begin{equation}\label{eq:row-sum}
\varepsilon(n-m-d_i^{in}) + \delta(n-1 - d_i^{in}) >0.
\end{equation}
Solving for $d_i^{in}$, we see that this inequality is satisfied when $d_i^{in} < \frac{\varepsilon(n-m) + \delta(n-1)}{\varepsilon+\delta}$.\\
 Next since
\begin{eqnarray*}
\frac{\varepsilon(n-m) + \delta(n-1)}{\varepsilon+\delta} & =& \frac{(\varepsilon+\delta)(n-m) + \delta(m-1)}{\varepsilon+\delta}\\
&=& n-m + \frac{\delta(m-1)}{\varepsilon+\delta}\\
&>& n-m + \frac{\delta(m-1)}{2\delta}\\
&=& n-\frac{m+1}{2},
\end{eqnarray*}
we see that whenever $d_i^{in} \leq n-\frac{m+1}{2}$, the inequality~\eqref{eq:row-sum} is satisfied.  Thus, by Theorem~\ref{thm:C-W} (Collatz-Wielandt), $\lambda_{\max} > n-(n-m)\varepsilon,$ and so $I-W$ must have a negative eigenvalue.  Therefore, $G$ is not a stable motif.  

To prove the last statement of the theorem, suppose that the skeleton $\Ghat$ has maximum in-degree $d_{\max}^{in}(\Ghat) <\frac{m}{2},$ so that $\widehat d_j^{in} \leq \frac{m-1}{2}$ for every node $j$ in the skeleton graph.  Then in the composite graph $G$, for every node in component $G_j$, there are at least $m-\frac{m-1}{2} = \frac{m+1}{2}$ components that the node does \underline{not} receive from in $G$.  Since every component is nonempty, this guarantees that each node $i$ in $G_j$ has at least $\frac{m+1}{2}$ nodes in $G$ that it does not receive from, and so $d_i^{in} \leq n- \frac{m+1}{2}$.  Applying the first part of the theorem we then see that $G$ is not a stable motif.  
\end{proof}

As an immediate corollary, we see that when the skeleton $\Ghat$ is an independent set or a cycle, so that $G$ is a disjoint union or cyclic union respectively, then $G$ is not a stable motif.

\begin{corollary}\label{cor:cyclic-union-unstable}
Let $G$ be a disjoint union or a cyclic union of cliques.  Then $G$ is not a stable motif.
\end{corollary}

%
%

\section{Stable motifs up to size 4 and proof of Theorem~\ref{thm:n4}}\label{sec:n4}

In this section, we prove Theorem~\ref{thm:n4} showing that for $\sigma$ of size at most 4, $\sigma$ is a stable motif if and only if it is a clique, and thus it can only support a stable fixed point in a larger network if it is a target-free clique.  For the proof, we analyze all permitted motifs up through size 4 and show that each non-clique cannot be a stable motif by applying key results from each of the earlier sections.  To perform this analysis, we must first generate a comprehensive list of these permitted motifs; the following lemma is key to efficiently generating this list.

\begin{lemma}\label{lemma:parameter-independent}
Let $\sigma$ have size $|\sigma| \leq 4$ and suppose $\sigma$ is a permitted motif of a CTLN $W=W(G, \varepsilon, \delta)$ for some legal choice of $\varepsilon$ and $\delta$.  Then $\sigma$ is a permitted motif for all $\varepsilon$ and $\delta$ in the legal range.  Furthermore, $\idx(\sigma)$ is constant across all $\varepsilon$ and $\delta$ in the legal range.
\end{lemma}
\begin{proof}
In \cite[Theorem 6]{fp-paper}, it was shown that the collection of fixed point supports, $\FP(W)$, is parameter independent for all graphs up to size 4.  Thus when $|\sigma| \leq 4$, if $\sigma$ is a permitted motif for a particular choice of $\varepsilon$ and $\delta$, then $\sigma$ is a permitted motif for all $\varepsilon$ and $\delta$ in the legal range.  

Furthermore, the proof of \cite[Theorem 6]{fp-paper} established that all permitted motifs up to size 3 have parameter-independent survival rules and have index that is constant across all $\varepsilon$ and $\delta$ in the legal range.  Since the index of a permitted motif of size 4 can be determined in a parameter-independent way via Theorem~\ref{thm:parity} (parity) from the sum of the indices of the surviving permitted motifs contained in it, it follows that $\idx(\sigma)$ is constant across all $\varepsilon$ and $\delta$ in the legal range for $\sigma$ of size 4 as well.
\end{proof}

It was shown in \cite{fp-paper} that there are permitted motifs of size 4 that have parameter-dependent survival in larger networks, and thus there are graphs of size 5 that are only permitted motifs for $\varepsilon$ and $\delta$ within a subset of the legal range (see Appendix A.2 in \cite{fp-paper}).  Thus, Lemma~\ref{lemma:parameter-independent} does not extend to larger $n$.

\begin{figure}[!ht]
\hspace{.5in} \includegraphics[width=.85\textwidth]{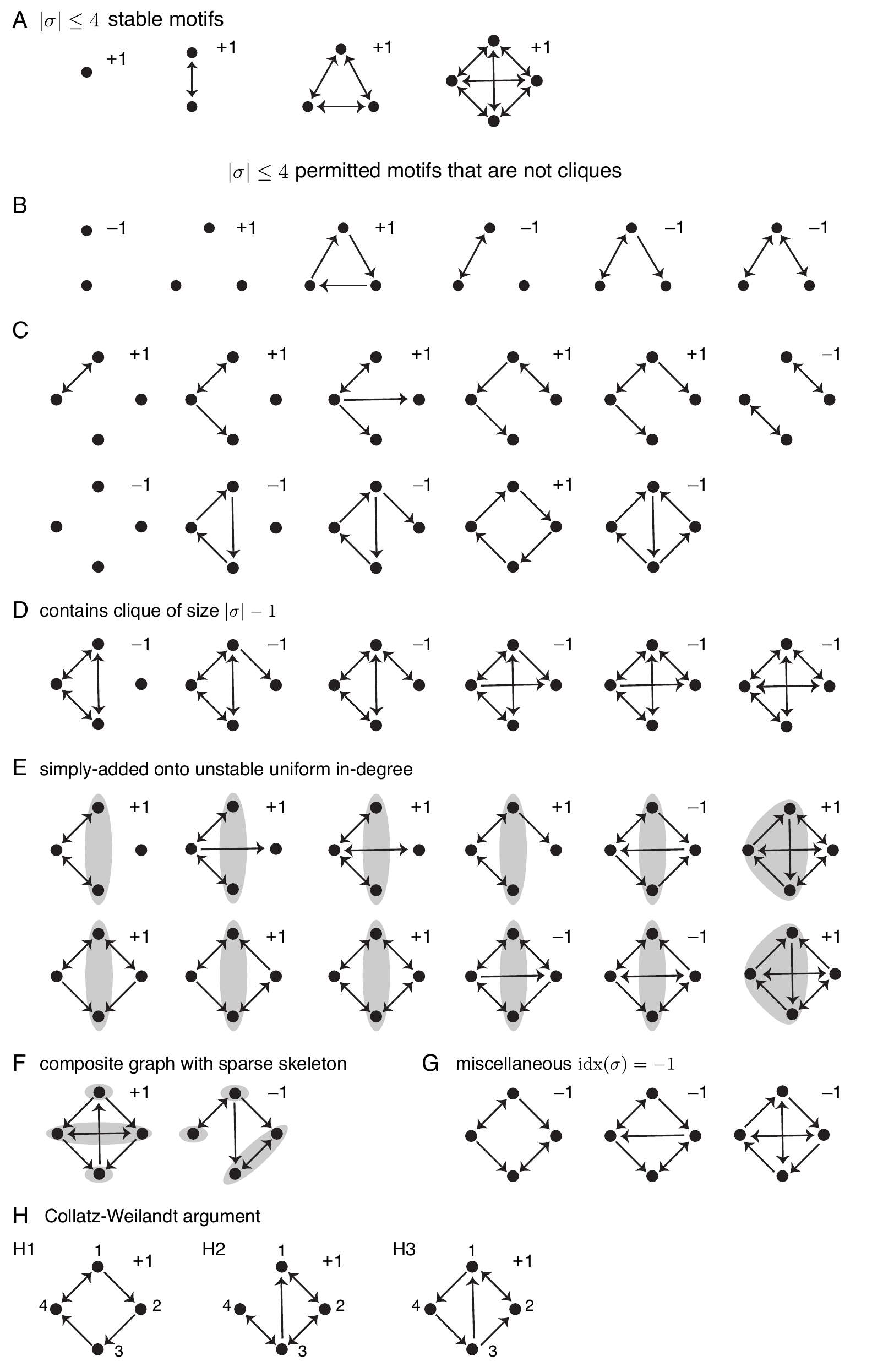}
\begin{center}
\vspace{-.1in}
\caption{All permitted motifs with $|\sigma|\leq 4$ annotated with their index on the top right of the graph.}
\label{fig:n34-permitted}
\end{center}
\vspace{-.1in}
\end{figure}

Figure~\ref{fig:n34-permitted} shows all the permitted motifs of size $|\sigma| \leq 4$ together with their index.  These were identified computationally by finding all graphs that have a full-support fixed point for $\varepsilon=0.25$ and $\delta=0.5$, which Lemma~\ref{lemma:parameter-independent} guarantees is sufficient to generate a complete, parameter-independent list.  It is worth noting that while all the indices were determined computationally, many of them could have actually been determined directly via Lemma~\ref{lemma:alternation-ufd} since they contain a uniform in-degree subgraph of size 1 less.

As an aside, one interesting observation from Figure~\ref{fig:n34-permitted} is that whenever a graph is permitted, its complement is also permitted.  Furthermore, for graphs of size 3, the index of a graph and its complement are the same, whereas for size 4, complementing flips the sign of the index.  It turns out that the pattern of a graph being permitted if and only if its complement is permitted does not hold for size 5 however (this breaks down in many of the cases when the graph contains a subgraph of size 4 that has parameter-dependent survival).  

We now turn to the proof of Theorem~\ref{thm:n4}, in which we will show that the only graphs in Figure~\ref{fig:n34-permitted} that are stable motifs are precisely the cliques (shown in Panel A).  To aid the proof, the graphs are organized in the figure by size and then by which results will be used to show they are not stable motifs.

\FloatBarrier

\begin{proof}[{\bf Proof of Theorem~\ref{thm:n4}}]
Panel B of Figure~\ref{fig:n34-permitted} shows all the permitted motifs of size $|\sigma|\leq 3$ that are not cliques.  
Observe that all these graphs have maximum in-degree $d_{\max}^{in} \leq 1 <\frac{|\sigma|}{2}$ and/or contain a clique of size $|\sigma|-1=2$.  Thus, by Theorems~\ref{thm:sparse-G} and \ref{thm:near-cliques}, none of these graphs is a stable motif.  

Next consider the permitted motifs of size 4 that are not cliques.  All the graphs in panel C of Figure~\ref{fig:n34-permitted} have maximum in-degree $d_{\max}^{in} \leq 1 < \frac{|\sigma|}{2}$ and/or are oriented, and so by Theorems~\ref{thm:sparse-G} and~\ref{thm:oriented} are not stable motifs.  The graphs in panel D all contain a clique of size $|\sigma|-1=3$, and thus by Theorem~\ref{thm:near-cliques} are not stable motifs.

For panel E of Figure~\ref{fig:n34-permitted}, observe that every graph has a simply-added split $\sigma= \tau\; \dot\cup\; \omega$ with $\omega$ simply-added onto a $\tau$ (the nodes in the gray shaded region) that is uniform in-degree.  Since $\tau$ is either an independent set or a $3$-cycle, and neither of these is a stable motif, Corollary~\ref{cor:simply-added-unstable} guarantees that $\sigma$ also is not a stable motif.  

The two graphs in panel F can be decomposed as composite graphs where each component is a clique.  The skeleton of the first composite graph is a $3$-cycle, while the skeleton of the second graph is a $2$-clique with an outgoing edge.  Both these skeletons have maximum in-degree 1, and thus by Theorem~\ref{thm:composite-unstable}, neither composite graph is a stable motif.  

The graphs in panel G do not fit into any of the families characterized by earlier results, but they all have index $-1$, and thus cannot be stable motifs.  

Finally, for the graphs in panel H of Figure~\ref{fig:n34-permitted}, we must appeal to more detailed arguments about the eigenvalues to show they are not stable motifs.  Specifically, we will use Theorem~\ref{thm:C-W} (Collatz-Wielandt) to show that these graphs have a maximum eigenvalue (or sum of eigenvalues) that is larger than the trace, thus forcing $I-W_\sigma$ to have a negative eigenvalue, and guaranteeing that $\sigma$ is not a stable motif. 

\begin{itemize}
\item[(H1)] For the graph in H1, 
\[I-W_\sigma = \begin{bmatrix}
  1 & 1+\delta  & 1+\delta & 1-\varepsilon \\
  1-\varepsilon & 1 & 1+\delta & 1+\delta\\
  1+\delta & 1-\varepsilon & 1 & 1+\delta\\
  1-\varepsilon & 1+\delta & 1-\varepsilon & 1
  \end{bmatrix}.
  \quad \quad
\]

Consider $x = [1,\ 1,\ 1,\ 1-\delta/2 ]^{\operatorname{T}}$, then 

$[(I-W_\sigma)x]_{1} = 3+2\delta +1 -\varepsilon-\delta/2+\varepsilon\delta/2 > 4$.\\
$[(I-W_\sigma)x]_{2} = [(I-W_\sigma)x]_{3} = 4+\delta/2-\delta^2/2+\delta-\varepsilon > 4$.\\
$[(I-W_\sigma)x]_{4} = 4+\delta/2-2\varepsilon > 4-2\varepsilon > 4(1-\delta/2)$.

Thus, by Theorem~\ref{thm:C-W} (Collatz-Wielandt), $\lambda_{\max} \geq \displaystyle\min_{i \in \sigma} \frac{[(I-W_\sigma)x]_i}{x_i} > 4$.  Since $\lambda_{\max}$ exceeds $\Tr(I-W_\sigma)=4$, $\sigma$ is not a stable motif.  

\item[(H2)] First observe that this graph has a simply-added split where $\{3,4\}$ is simply-added onto the clique $\{1,2\}$.  Thus, by Lemma~\ref{lemma:simply-added-evals}, $\sigma$ inherits the eigenvalue $\varepsilon$ from the clique.  Therefore, we need only show that $\lambda_{\max}$ of $I-W_\sigma$ exceeds $4-\varepsilon$ to guarantee that a sum of eigenvalues exceeds $\Tr(I-W_\sigma)=4$.  Observe that 
\[I-W_\sigma = \begin{bmatrix}
  1 & 1-\varepsilon & 1-\varepsilon  & 1+\delta\\
  1-\varepsilon & 1 & 1-\varepsilon  & 1+\delta\\
  1+\delta  & 1-\varepsilon & 1 & 1-\varepsilon\\
  1+\delta  & 1+\delta & 1-\varepsilon & 1
  \end{bmatrix}
    \quad \quad
\]
and consider $x = [1,\ 1,\ 1,\ 1+\varepsilon/2]^{\operatorname{T}}$. Then

$[(I-W_\sigma)x]_{1} = [(I-W_\sigma)x]_{2} =  3-2\varepsilon + 1+\delta+(1+\delta)\varepsilon/2> 4-2\varepsilon+\delta>4-\varepsilon$.\\
$[(I-W_\sigma)x]_{3} =  3 + \delta - \varepsilon  +1 -\varepsilon + (1-\varepsilon)\varepsilon/2 > 4-2\varepsilon + \delta > 4-\varepsilon$.\\
$[(I-W_\sigma)x]_{4} = 4+2\delta-\varepsilon/2 > 4+\varepsilon > 4+\varepsilon - \varepsilon^2/2= (4-\varepsilon)(1+\varepsilon/2)$.\\
Thus by the Collatz-Wielandt Theorem, $\lambda_{\max} > 4-\varepsilon$, and $\lambda_{\max} + \varepsilon > \Tr(I-W_\sigma)$.  Therefore, $\sigma$ is not a stable motif.

\item[(H3)] As with the previous graph, there is a simply-added split where $\{3,4\}$ is simply-added onto the clique $\{1,2\}$, and so $\sigma$ inherits the eigenvalue $\varepsilon$ from the clique. We will again use Theorem~\ref{thm:C-W} (Collatz-Wielandt) to show that $\lambda_{\max} > 4-\varepsilon$.  Observe that 

\[I-W_\sigma = \begin{bmatrix}
  1 & 1-\varepsilon & 1-\varepsilon & 1+\delta\\
  1-\varepsilon & 1 & 1-\varepsilon & 1+\delta\\
  1+\delta & 1+\delta & 1 & 1-\varepsilon\\
  1-\varepsilon & 1+\delta & 1+\delta & 1
  \end{bmatrix}
    \quad \quad
\]
and again consider $x = [1,\ 1,\ 1,\ 1+\varepsilon/2]^{\operatorname{T}}$. Then 

$[(I-W_\sigma)x]_{1} = [(I-W_\sigma)x]_{2} =  3-2\varepsilon + 1+\delta+(1+\delta)\varepsilon/2> 4-2\varepsilon+\delta>4-\varepsilon$.\\
$[(I-W_\sigma)x]_{3} =  3+ 2\delta +1 -\varepsilon + (1-\varepsilon)\varepsilon/2 =4- \varepsilon +2\delta + (1-\varepsilon)\varepsilon/2> 4-\varepsilon$.\\
$[(I-W_\sigma)x]_{4} = 4+2\delta-\varepsilon/2 > 4+\varepsilon > 4+\varepsilon - \varepsilon^2/2= (4-\varepsilon)(1+\varepsilon/2)$.\\
Thus, $\lambda_{\max} > 4-\varepsilon$, and so $I-W_\sigma$ has a sum of eigenvalues larger than its trace, forcing a negative eigenvalue.  Hence $\sigma$ is not a stable motif.

\end{itemize}

\noindent Therefore every permitted motif up to size 4 that is not a clique is unstable, and so if $|\sigma|\leq 4$, then $\sigma$ supports a stable fixed point if and only if $\sigma$ is a target-free clique.  
\end{proof}

\FloatBarrier

\section{Enumerating target-free cliques and proof of Theorem~\ref{thm:maximal-cliques}} \label{sec:maximal-cliques}
Thus far we have seen that for certain parameter regimes, the only stable motifs are cliques, and we have also ruled out a large variety of graphs from being stable motifs for every choice of parameters within the legal range.  This provides strong evidence for the conjecture that the only stable motifs are cliques.  Furthermore, since cliques only survive to yield fixed points of the full network when they are target-free, a critical step to understanding the collection of stable fixed points of a CTLN is to be able to find and count the target-free cliques of its underlying directed graph.  

In this section, we prove a correspondence between target-free cliques of a directed graph and maximal cliques of an undirected graph.  Counting and enumerating maximal cliques of undirected graphs has garnered significant attention in extremal graph theory, and thus this correspondence will enable us to easily import results from this field to yield information about the target-free cliques of a directed graph, and thus about the stable fixed points of corresponding CTLNs.

Given a directed graph $G$, we can construct a corresponding undirected graph $\Gtil$ by making all bidirectional edges in $G$ into undirected edges in $\Gtil$ and dropping all other edges, i.e.\ $(i,j)$ is an undirected edge in $\Gtil$ if and only if $i \leftrightarrow j$ in $G$.  Then every target-free clique of $G$ corresponds to a maximal clique in $\Gtil$ (although $\Gtil$ may have additional maximal cliques as well that were originally targeted in $G$).  Thus, the number of target-free cliques in $G$ is at most the number of maximal cliques in $\Gtil$.  In fact, the following lemma shows that the maximum number of target-free cliques in directed graphs actually equals the maximum number of maximal cliques in undirected graphs.

\begin{lemma}\label{lemma:num-maximal-cliques}
The maximum number of target-free cliques in a directed graph equals the maximum number of maximal cliques in an undirected graph.  
\end{lemma}
\begin{proof}
From the above construction, we see that the maximum number of target-free cliques in a directed graph is less than or equal to the maximum number of maximal cliques in an undirected graph.  To see the reverse inequality, observe that given any undirected graph, there is a canonical corresponding directed graph obtained by replacing each undirected edge with a bidirectional edge.  In this case, all the maximal cliques of the undirected graph yield target-free cliques in the corresponding directed graph.  Thus, the maximum number of target-free cliques in a directed graph is greater than or equal to the maximum number of maximal cliques in an undirected graph, and so equality holds.  
\end{proof}

Lemma~\ref{lemma:num-maximal-cliques} allows us to immediately translate upper bounds on the number of maximal cliques in an undirected graph into upper bounds on the number of target-free cliques in directed graphs, and thus into upper bounds on the number of stable fixed points in CTLNs in parameter regimes where the conjecture holds.  For example, applying an upper bound from Moon and Moser \cite{moon-moser} immediately yields Theorem~\ref{thm:maximal-cliques}, restated below.\\

\noindent {\bf Theorem~\ref{thm:maximal-cliques}.}
\emph{The maximum number of target-free cliques in a directed graph of size $n$ is
$$ \text{max } \# \text{ of target-free cliques} = \left\{\begin{array}{cl}3^{n/3} & \textrm{if } n \equiv 0 \pmod 3 \\ 
4\cdot 3^{\lfloor n/3 \rfloor -1} &  \textrm{if } n \equiv 1 \pmod 3 \\
2\cdot 3^{\lfloor n/3 \rfloor} &  \textrm{if } n \equiv 2 \pmod 3 .
\end{array}\right. \quad \quad $$
}

Lemma~\ref{lemma:num-maximal-cliques} allows us to import a number of other results from the extremal graph theory literature as well.  For example, Hedman gives a tighter upper bound on the number of maximal cliques whenever the size of the largest clique is at most $n/2$ \cite{hedman}.  Moon and Moser also give an upper bound on the number of different sizes of maximal cliques \cite{moon-moser}, while Spencer gives a lower bound on this for particularly large $n$ \cite{spencer}, which improves on a previous lower bound of Erd\"os \cite{erdos}.  

Additionally, the construction of an undirected $\Gtil$ from a directed graph $G$ enables us to apply algorithms for finding maximal cliques in order to enumerate all the target-free cliques of $G$.  Specifically, the list of maximal cliques of $\Gtil$ gives all the maximal cliques of $G$, and it is straightforward to check which of these candidate cliques are in fact target-free in $G$.  Thus, target-free cliques can be easily found using algorithms for finding maximal cliques such as those from Bron and Kerbosch \cite{bron} (which has worst case running time of $O(3^{n/3})$), based on a branch-and-bound technique, or that of Tomita et al.\ \cite{tomita}, based on a depth-first search algorithm with pruning.  

\bigskip
\noindent{\bf Acknowledgments.} All three authors were supported by NIH R01 EB022862.  The first author was also supported by NSF DMS-1516881 and NSF DMS-1951165, while the third author was also supported by NSF DMS-1951599.

\bibliographystyle{unsrt}
\bibliography{CTLN-refs}
\end{document}

%% file: packages-v2.tex
\usepackage{amsmath}
\usepackage{amsthm}
\usepackage{amsfonts}
\usepackage{amssymb}
\usepackage{latexsym} 
\usepackage{epsfig}
\usepackage{graphicx}

\usepackage{calc}  
\usepackage[matrix,tips,graph,curve,web]{xy}

\makeatletter

\def\@maketitle{%
  \newpage
  \null
  \let \footnote \thanks
    {\normalfont\sffamily\bfseries\Large\noindent\@title \par}%
    \vskip 1em%
    {\normalfont\sffamily 
        \noindent
        \@author
        \par}
  \par
  \vskip 4em}
\def\@seccntformat#1{\csname the#1\endcsname{.\ }}
\renewcommand\section{\@startsection {section}{1}{\z@}%
                                   {-3.0ex \@plus -1ex \@minus -.2ex}%
                                   {1.5ex \@plus.2ex}%
                                   {\normalfont\large\bfseries}}
\renewcommand\subsection{\@startsection{subsection}{2}{\z@}%
                                     {-2.75ex\@plus -1ex \@minus -.2ex}%
                                     {1.5ex \@plus .2ex}%
                                   {\normalfont\large}}
\def\fnum@figure{\normalfont\footnotesize\figurename~\thefigure}

\renewcommand\tableofcontents{%
    \section*{\contentsname
        \@mkboth{%
           \MakeUppercase\contentsname}{\MakeUppercase\contentsname}}%
    \@starttoc{toc}%
    }
\renewcommand*\l@part[2]{%
  \ifnum \c@tocdepth >-2\relax
    \addpenalty\@secpenalty
    \addvspace{2.25em \@plus\p@}%
    \begingroup
      \setlength\@tempdima{3em}%
      \parindent \z@ \rightskip \@pnumwidth
      \parfillskip -\@pnumwidth
      {\leavevmode
       \large \bfseries #1\hfil \hb@xt@\@pnumwidth{\hss #2}}\par
       \nobreak
       \if@compatibility
         \global\@nobreaktrue
         \everypar{\global\@nobreakfalse\everypar{}}%
      \fi
    \endgroup
  \fi}
\renewcommand*\l@section[2]{%
  \ifnum \c@tocdepth >\z@
    \addpenalty\@secpenalty
    \addvspace{1.0em \@plus\p@}%
    \setlength\@tempdima{1.5em}%
    \begingroup
      \parindent \z@ \rightskip \@pnumwidth
      \parfillskip -\@pnumwidth
      \leavevmode \sffamily\bfseries
      \advance\leftskip\@tempdima
      \hskip -\leftskip
      #1\nobreak\hfil \nobreak\hb@xt@\@pnumwidth{\hss #2}\par
    \endgroup
  \fi}
\renewcommand*\l@subsection{\sffamily\@dottedtocline{2}{1.5em}{2.3em}}
\renewcommand*\l@subsubsection{\@dottedtocline{3}{3.8em}{3.2em}}
\renewcommand*\l@paragraph{\@dottedtocline{4}{7.0em}{4.1em}}
\renewcommand*\l@subparagraph{\@dottedtocline{5}{10em}{5em}}
\makeatother


\theoremstyle{plain}
\newtheorem{theorem}{Theorem}[section]
\newtheorem{corollary}[theorem]{Corollary}
\newtheorem{lemma}[theorem]{Lemma}
\newtheorem{proposition}[theorem]{Proposition}
\newtheorem{conjecture}[theorem]{Conjecture}

\theoremstyle{definition}
\newtheorem{definition}[theorem]{Definition}

\newtheorem{remark}[theorem]{Remark}

  {\begin{list}{}%
    {%
    \settowidth{\labelwidth}{#1}%
    \setlength{\itemindent}{0pt}%
    \setlength{\labelsep}{1em}%
    \setlength{\leftmargin}{\labelwidth+\parindent+\labelsep}%
    \setlength{\itemsep}{0pt}%
    \setlength{\parsep}{.6ex}}}%
  {\end{list}}

  {\begin{list}{}%
    {%
    \settowidth{\labelwidth}{#1}%
    \setlength{\itemindent}{0pt}%
    \setlength{\labelsep}{1em}%
    \setlength{\leftmargin}{\labelwidth+\labelsep}%
    \setlength{\itemsep}{0pt}%
    \setlength{\parsep}{.6ex}}}%
  {\end{list}}

\makeatletter
\newenvironment{subequations*}{
  \begingroup 
  \let\protect\@nx
  \edef\@tempa{\def\@nx\theparentequation{\theequation}}%
  \@xp\endgroup\@tempa
  \setcounter{parentequation}{\value{equation}}%
  \setcounter{equation}{0}%
  \def\theequation{\theparentequation\alph{equation}}%
  \ignorespaces
}{%
  \setcounter{equation}{\value{parentequation}}%
  \global\@ignoretrue
}

\newcommand{\sgn}{\rm sgn\,}

\renewcommand\det{{\rm det\,}}

\def\d/{/\mspace{-6.0mu}/}